\documentclass[12pt,showpacs]{revtex4}

\usepackage[french,english]{babel}
\usepackage[T1]{fontenc}
\usepackage{times}

\usepackage{dcolumn,amsmath}
\usepackage{amsfonts}
\usepackage{amsthm}
\usepackage{mathrsfs}
\usepackage{amssymb}
\usepackage{epsfig}
\usepackage{psfrag}

\newcommand{\PHI}{\mathbf{\Phi}}
\newcommand{\EXP}[1]{\mathrm{e}^{#1}}

\newcommand{\imat}{\mathrm{i}}

\newcommand{\be}{\begin{equation}}
\newcommand{\ee}{\end{equation}}
\newcommand{\X}{X}
\newcommand{\D}{\hat{{\cal D}}}

\begin{document}
\selectlanguage{english}

\title{Stability Analysis of The Twisted Superconducting Semilocal Strings}

\author{Julien Garaud and Mikhail S. Volkov}

\vspace{1 cm}

 \affiliation{ {Laboratoire de Math\'{e}matiques et Physique Th\'{e}orique
CNRS-UMR 6083, \\ Universit\'{e} de Tours,
Parc de Grandmont, 37200 Tours, FRANCE}
}

\vspace{1 cm}

\begin{abstract}
We study the stability properties of the twisted vortex solutions in the
semilocal Abelian Higgs model with a global $\mathbf{SU}(2)$ invariance. This model  
can be viewed as the Weinberg-Salam theory in the limit where the
non-Abelian gauge field decouples, or as a two component Ginzburg-Landau theory. 
The twisted vortices are characterized
by a constant global current ${\cal I}$, and for  ${\cal I}\to 0$ they
reduce to the semilocal strings, that is to the Abrikosov-Nielsen-Olesen vortices
embedded into the semilocal model. Solutions with ${\cal I}\neq 0$ 
are more complex and, in particular, they are
{\it less energetic} than the semilocal strings, which makes one hope that they
could have better stability properties. 
We consider  the generic field fluctuations around the twisted vortex 
within the linear perturbation theory and 
apply the Jacobi criterion to test the existence of the negative modes in the 
spectrum of the fluctuation operator. 
We find that twisted vortices do not have the  
homogeneous instability known for the semilocal strings,
neither do they have inhomogeneous instabilities whose wavelength
is less than a certain critical value. This implies that short enough vortex pieces
are perturbatively stable and suggests that small vortex loops
could perhaps be  stable as well. For longer wavelength perturbations there is 
exactly one negative mode in the spectrum whose growth 
entails a segmentation 
of the uniform vortex into a non-uniform, `sausage like' structure.
This instability is qualitatively similar to the  
 hydrodynamical Plateau-Rayleigh instability of a water jet   
or to the Gregory-Laflamme instability of black strings 
in the theory of gravity in higher dimensions. 
\end{abstract}


\maketitle

\newpage

\section{Introduction}

Vortices have been the subject of intense studies ever since their discovery by 
Abrikosov in the context of the Ginzburg-Landau theory 
of superconductivity and  by Nielsen and Olesen in  
relativistic field theory \cite{ANO}. They find numerous applications in many domains of 
physics ranging from high energy physics and cosmology \cite{vil} to 
various branches of condensed matter physics, such as superconductivity
\cite{sigrist}
and superfluidity  \cite{vol} models. 
In these applications the vortices are most often considered within the original 
model of Abrikosov, Nielsen and Olesen (ANO) \cite{ANO}, which means as solutions 
of the Abelian Higgs model containing an Abelian vector $A_\mu$ coupled to a complex 
scalar $\Phi$. 
However, vortices can 
also exist in others, more general field theory models.  

A particular field theory containing an Abelian vector $A_\mu$ 
and two complex scalars $\Phi_1$ and $\Phi_2$ with a global 
SU(2) invariance, dubbed semilocal model, has been extensively discussed recently 
\cite{achuc}. This model can be considered as a `minimal' generalization of the original 
ANO theory with the following interesting properties. 
First, it can be viewed as the special
limit of the electroweak theory of Weinberg and Salam in 
which the weak mixing angle is $\pi/2$
and the non-Abelian gauge field decouples \cite{achuc}. 
Secondly, it describes states with a multicomponent order parameter 
in condensed matter physics, as for example  
it the two band superconductivity models or in superfluid  systems
\cite{baba}. Finally, it can be regarded as sector
of supersymmetric field theories \cite{Shifman} or as part 
of one of the Grand Unification
Theories that could presumably describe 
cosmic strings  in the early Universe \cite{Witten}. 
All this implies that the semilocal model is likely 
to have interesting physical application, which has  
inspired considerable interest towards 
this theory and  its solutions. 

The original ANO vortex can  be embedded into the semilocal theory, 
since identifying its field $A_\mu$ with that of the semilocal model and setting also 
$\Phi_1=\Phi$ and $\Phi_2=0$ solves the field equations of the theory \cite{vacha}. 
Such an embedded vortex is called semilocal string  and it behaves identically to the 
original vortex as long as $\Phi_2=0$, but its properties can become quite different as 
soon as the scalar $\Phi_2$ is excited \cite{achuc}. 
For example, although the original 
ANO vortex is stable, fluctuations of $\Phi_2$ render its embedded version unstable
in the parameter region where the Higgs boson mass is larger than the vector boson mass
\cite{Hind}.  

Not all vortices in the semilocal theory should necessarily be of 
the embedded ANO type. It has been known for some time that more general vortex solutions
(sometimes called  `skyrmions') exist in the theory  
in the special limit where the Higgs boson mass is equal to the vector boson mass
\cite{Hind}, \cite{Gibbons}.  For these solutions  $\Phi_1$ and $\Phi_2$ are 
both non-vanishing. Quite recently vortex solutions with a similar property have been 
constructed within 
the whole  parameter region of the theory in which the Higgs boson mass is larger than the 
vector boson mass \cite{SL}. 
They are characterized by the twist: a $z$-dependent relative phase 
$\exp(iKz)$ between $\Phi_1$ and $\Phi_2$.
The twist gives rise to a constant global 
current along the vortex, whose value, ${\cal I}$, is a parameter of the solutions.
For this reason the solutions are called twisted superconducting semilocal strings,
or twisted vortices for short. 
For ${\cal I}\to 0$ they reduce to the semilocal strings but for 
${\cal I}\neq 0$ they are physically different. In particular, it turns out that 
solutions with ${\cal I}\neq 0$ are {\it less energetic} 
 as compared to the semilocal strings.  This suggests that, since they 
are energetically favoured,  it is predominantly the twisted vortices 
 and not the semilocal strings 
which should be created   in physical processes leading to vortex
formation.  Therefore, 
as far as physical applications is concerned, 
twisted vortices could be more important than the semilocal strings.
However, since they  exist precisely  in the same parameter 
region where the latter are unstable, the twisted vortices may well 
be also unstable, which would somewhat delimit their importance. 
On the other hand, one may argue that   they should have   
better stability properties than the semilocal strings, 
since they are less energetic.

Motivated by this, we study in this paper the stability of the
twisted vortices within the linear perturbation theory by analysing 
the linearized equations for fluctuations around the twisted vortex
background. Decomposing the perturbations into a sum over Fourier modes 
proportional to  
$\exp\{i(\omega t+\kappa z + m\varphi)\}$,
the variables in the fluctuation equations separate and the equations reduce to 
a Schrodinger type spectral problem with the eigenvalue $\omega^2$.
If this problem admits eigenstates  with $\omega^2<0$ then the perturbations
will be growing in time
and so the background will be unstable. In order to find out whether 
such negative modes exist, we apply the Jacobi criterion
\cite{jack}, which only requires the knowledge of the $\omega=0$ solutions
of the fluctuation equations.  

We arrive at the following conclusions. 
For ${\cal I}\neq 0$ the perturbation equations do not admit negative mode solutions 
independent of $z$. This means that the twisted vortices do not have 
the homogeneous {instability} 
known for the semilocal strings \cite{Hind}. Moreover, 
considering inhomogeneous $z$-dependent perturbations of 
the fundamental twisted vortex, we find that the corresponding eigenvalues are 
always positive, apart from those corresponding to the   
modes with $m=0$ and with the 
wavelength 
\be                                 \label{0}
\lambda>\lambda_{\rm min}=\frac{\pi}{|K|},
\ee
where $K$ is the twist parameter of the vortex. 
It follows then that short vortex pieces obtained by imposing the periodic 
boundary conditions on the infinite vortex will be perturbatively stable 
if their length, $L$, is less then $\lambda_{\rm min}$. 
If $L>\lambda_{\rm min}$  then the vortex will have enough room to accommodate 
inhomogeneous instability modes whose growth will lead to its fragmentation 
into a non-uniform, `sausage like' structure characterized by 
zones of charge accumulation and by an inhomogeneous 
current density. One needs to go beyond the linearized approximation
in order to fully understand the development of this instability. 
For higher winding number vortices 
we find additional instabilities corresponding to their splitting into 
vortices of lower winding number. 

Our general conclusion is that the twisted vortices indeed have better 
stability properties
than the semilocal strings since, unlike the latter, they can be 
stabilized by imposing
periodic boundary conditions. This suggests 
that small stationary vortex loops, 
if exist, might be stable, which opens intriguing research perspectives
that may lead to interesting applications in various domains of physics ranging   
form condensed matter physics to cosmology.  

The rest of the paper is organized
as follows. In Sec.II we summarize the essential properties 
of the  twisted vortices. Sec.III contains the analysis of generic 
perturbations around the twisted
vortex background in the linearized approximation: the mode decomposition, 
separation of variables in the perturbation equations, gauge fixing, 
and the reduction to a Schrodinger type spectral problem.  
The existence of the negative modes in the spectrum is demonstrated  
with the use of the 
Jacobi criterion in Sec.IV, where these modes are also 
explicitly constructed and their dispersion 
relation is determined. The physical manifestations of the instability 
are discussed in Sec.V, where analogies from other branches of physics
are also discussed.  
We summarize our results
in Sec.VI, while the complete system of perturbation equations is 
presented in the Appendix. 

The issue of vortex perturbations   
is not so often considered in the literature. 
Although the basic stability
pattern for the ANO vortices was understood long ago \cite{BV},
the systematic stability analysis in this case was carried out only 
relatively recently \cite{Good}. In our analysis we essentially follow 
the approach of  \cite{Good}, although working in a different gauge.

\section{The Twisted Semilocal Vortices}

\subsection{The semilocal model}

The model considered here is obtained by replacing the complex scalar of the 
usual Abelian-Higgs model by a doublet of complex scalars, 
$\PHI^{\rm tr}=(\Phi_1,\Phi_2)$. This theory, the \textit{semilocal model}, is by construction 
invariant under the internal 
symmetry  $\mathbf{SU}(2)_{\rm global}\times \mathbf{U}(1)_{\rm local }$. 
It is worth noting that this semilocal model describes the 
bosonic sector of the electroweak theory 
in the limit where the weak mixing angles is $\pi/2$ and the non-Abelian 
gauge field decouples \cite{achuc}. 

The Lagrangian density in terms of suitably rescaled dimensionless coordinates and 
fields reads 
\begin{equation}\label{lagr}
  \mathcal{L}=-\frac{1}{4}F_{\mu\nu}F^{\mu\nu}+\left(D_\mu\PHI\right)^\dagger 
  D^\mu\PHI-\frac{\beta}{2}\left(\PHI^\dagger\PHI-1\right)^2,
\end{equation}
where $F_{\mu\nu}=\partial_\mu A_\nu-\partial_\nu A_\mu$ and 
$D_\mu\PHI = \partial_\mu\PHI -\imat A_\mu\PHI$, the 
spacetime metric signature being $(+,-,-,-)$. 
The spectrum of this theory consists of a massive vector boson whose
mass in the dimensionless units chosen is $m_v=\sqrt{2}$, of two massless
Goldstone scalars, and of a  Higgs boson with the mass 
$m_{\mbox{\tiny H}}=\sqrt{\beta}\,m_v$.  
The fields transform under 
$\mathbf{SU}(2)_{\rm global}\times \mathbf{U}(1)_{\rm local }$
as 
\begin{equation}                 \label{local}
A_\mu\rightarrow A_\mu +\partial_\mu\Lambda(x),\qquad
\PHI\rightarrow\mathbf{U}\PHI, 
\end{equation}    
with $\mathbf{U}=\EXP{\imat\Lambda(x)+\imat\theta_a\tau^a}$  
where $\tau_a$ are the Pauli matrices and $\theta^a$ are constant parameters. 
The Euler-Lagrange equations of motion obtained by varying the Lagrangian 
$(\ref{lagr})$ read 
\begin{align}\label{eqs}
  \partial^\mu F_{\mu\nu}&=\imat\{\left(D_\nu\Phi\right)^\dagger\Phi-\Phi^\dagger
  D_\nu\Phi \}, \notag \\
  D_\mu D^\mu\Phi&=-\beta\left(\Phi^\dagger\Phi-1\right)\Phi,
\end{align}
where the dagger stands for the hermitian conjugation.

The internal symmetries of the theory
give rise to several independently conserved Noether currents, one of which will be important 
in what follows,
\be                                    \label{cur}
J_\mu=\Re (i\Phi_2^\ast D_\mu\Phi_2).
\ee

\subsection{The twisted superconducting vortices}

Eqs.\eqref{eqs} admit stationary, cylindrically symmetric solutions of vortex type
\cite{SL}.  
For these solutions the fields are parametrized in 
cylindrical coordinates as
\begin{align}\label{ansatz1}
A_\mu d x^\mu=a_2(\rho)(\Omega\,dt+ K\,dz)+Na_1(\rho)d\varphi,~~
  \PHI =\left(\begin{array}{c}f_1(\rho)\EXP{\imat N\varphi} \\
    f_2(\rho)\EXP{\imat(\Omega t+M\varphi+Kz)} \end{array}\right),
\end{align}
where all functions of $\rho$ are real and $N,M$ are two integer winding numbers.
The two real parameters $\Omega$ and 
$K$ can be seen, respectively, as the relative rotation and twist between 
the two components of the scalar doublet. 
%The ansatz \eqref{ansatz1} preserves 
%its structure under Lorentz boosts in the $t,z$ plane
With this parametrization 
Eqs.\eqref{eqs} reduce to a set of four 
non-linear ordinary differential equations, 
\begin{align}\label{4ode}
  \frac{1}{\rho} \left( \rho a_2^\prime\right)^\prime 
  &= 2a_2f_1^2+2(a_2-1)f_2^2  ,\notag\\
  \rho\left(\frac{a_1^\prime}{\rho}\right)^\prime 
  &= 2f_1^2(a_1-1)+2f_2^2\left(a_1-\frac{M}{N}\right)  ,\notag\\
  \frac{1}{\rho}(\rho f_1^\prime)^\prime &=f_1\left(\frac{N^2(a_1-1)^2}
       {\rho^2}+(K^2-\Omega^2)a_2^2+\beta\left(f_1^2+f_2^2-1\right)\right),  \notag\\ 
       \frac{1}{\rho}(\rho f_2^\prime)^\prime &=f_2\left(\frac{(M-Na_1)^2}{\rho^2}
       +(K^2-\Omega^2)(a_2-1)^2+\beta\left(f_1^2+f_2^2-1\right)\right),		
\end{align}
with $\enskip ^\prime=\frac{d}{d\rho}$. One can consistently set in these equations 
$f_2=a_2=0$. The remaining two non-trivial equations for $f_1$, $a_1$ reduce then to 
the ANO system \cite{ANO}
and so the solutions will be the ANO vortices embedded into the semilocal theory. 
Such embedded solutions are sometimes called semilocal strings \cite{achuc}.  

\begin{figure}[ht]
\hbox to\linewidth{\hss%

\psfrag{rho}{$\rho$}
	\psfrag{D}{}
	\psfrag{f1}{$f_1$}
	\psfrag{a1}{$a_1$}
	\psfrag{f2}{$f_2$}
	\psfrag{a2}{$a_2$}
	\resizebox{8cm}{6cm}{\includegraphics{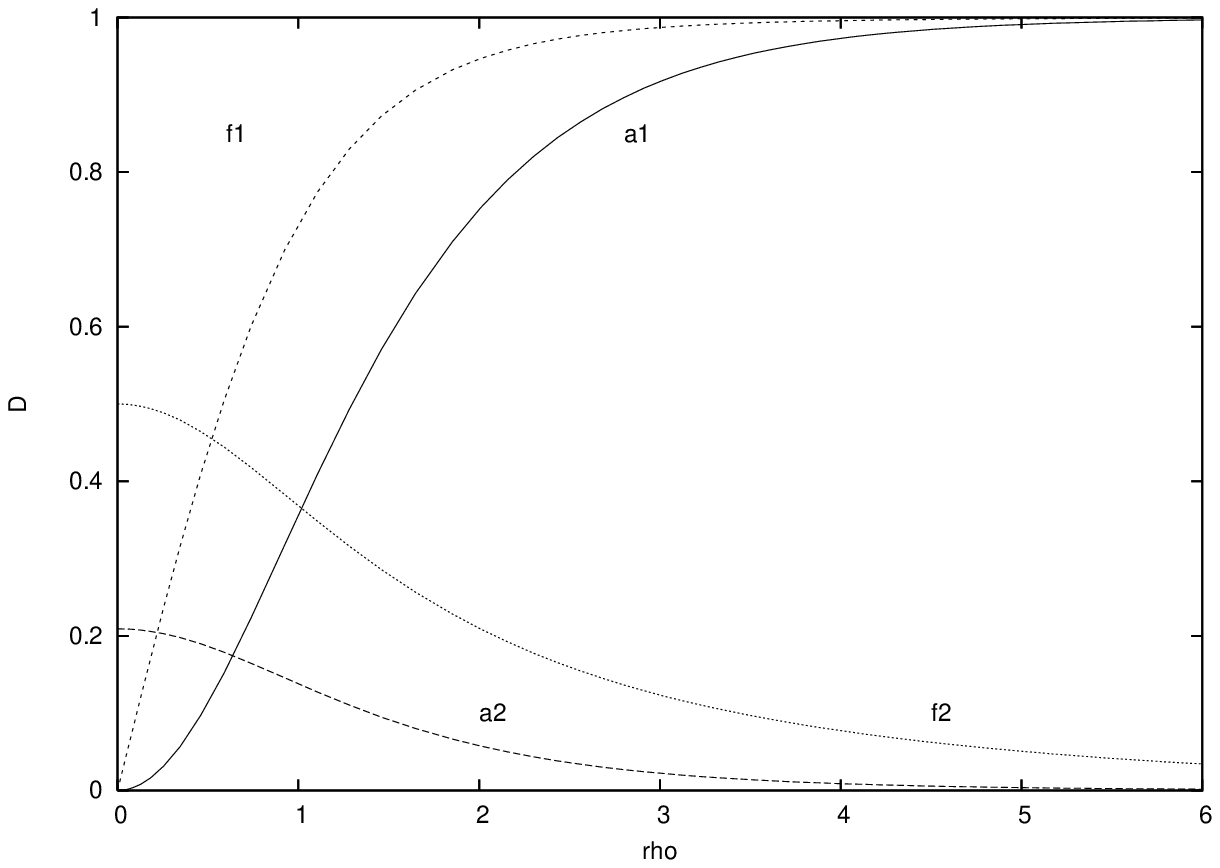}}
\hspace{5mm}%

	\psfrag{q}{$q$}
	\psfrag{E}{$E/4\pi-1$}
	\psfrag{I}{$|I|/2\pi$}
	\psfrag{K}{$K$}
	\resizebox{8cm}{6cm}{\includegraphics{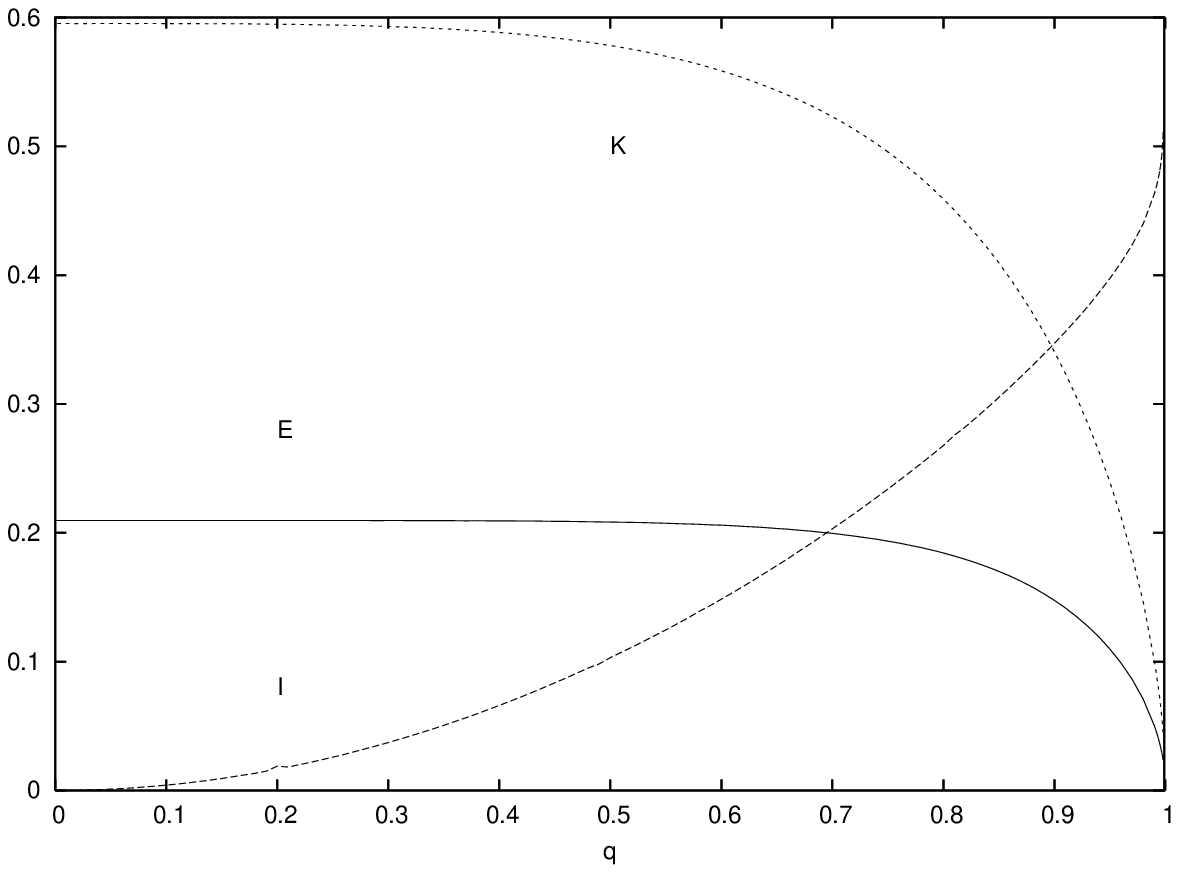}}
\hss}
	\caption{Left: profile functions for the twisted vortex solution with 
$\beta = 2$, $N= 1$ and $q=0.5$. Right: the restframe ($\Omega=0$)
vortex energy $E$, current ${\cal I}$, 
and twist 
$K$ 
against the condensate parameter $q$ for the $N=2$, $M=0$ twisted solutions
with $\beta=2$. 
}
\label{Fig1}
\end{figure}

In addition,  for $\beta>1$, 
Eqs.\eqref{4ode} possess also more general solutions, called twisted vortices, 
which have
 $f_2\neq 0$ and $a_2\neq 0$ \cite{SL}. 
Their numerical profiles (see Fig.\ref{Fig1}) show that 
$f_1,a_1$ behave qualitatively in the same way as for the ANO vortex, while 
$f_2$, $a_2$  develop non-zero condensate values in the vortex core and tend to zero for 
$\rho\to\infty$.  
For a given $\beta>1$ these solutions comprise a three parameter family labeled 
by $N=1,2,\ldots\,$, by $M=0,1,\ldots N-1$, and by a real parameter  
$q=\frac{1}{M!}f^{(M)}_2(0)$: the value of the $M$-th derivative of $f_2$ at the vortex center. 
In what follows we shall call $q$ condensate parameter and shall be mainly
considering solutions with $M=0$, so that $q=f_2(0)$. 
These parameters determine the value of the combination 
$
K^2-\Omega^2
$ which turns out to be positive for all twisted solutions. It is worth noting that 
the ansatz \eqref{ansatz1} preserves its structure under Lorentz boosts along the 
vortex axis -- if we assume that $(\Omega,K)$ transform as components of a 
spacetime vector. Since the Lorentz invariant norm of this vector, $K^2-\Omega^2$,
is positive, the vector is spacelike, and so one can boost away 
its temporal component by passing to the restframe where  
$\Omega=0$. On the other hand, the twist $K$
is an essential parameter that cannot be removed, which is why the vortices are
called  \textit{twisted}. 
Since the equations are invariant under $K\to -K$,
in what follows we can assume without loss of generality that $K>0$.

Associated to the twist there is a 
physical parameter:  the total current  \eqref{cur} through the vortex
cross section, 
\begin{equation}\label{current}
\mathcal{I}=\int J_z\, \text{d}^2x=\int\text{d}^2x(A_z-K)\Phi_2^\ast\Phi_2.
\end{equation}
This is zero for the embedded ANO solutions and non-zero for the twisted 
vortices.  
The current is a function of the condensate parameter $q$ (also of $N,M$). 
In the limit $q\to 0$ one has $f_2\to 0$, $a_2\to 0$, the current 
vanishes and the twisted vortices reduce to the semilocal strings. 
For $q\neq 0$ the current is non-zero and it becomes arbitrarily large 
when $q\to 1$ \cite{SL} (although this is not so visible in Fig.\ref{Fig1}). 
Interestingly, both the twist $K$ and the vortex energy per unit length,
$E=\int T^0_0 d^2x$,   
{\sl decrease} when the current increases (see Fig.\ref{Fig1}), for 
large currents the energy approaching the lower bound $2\pi|N|$ \cite{SL}.  
The twisted current carrying vortices are thus energetically favored when compared to 
the currentless semilocal strings, and so that 
one may expect that they could have better stability properties than the 
latter.

\section{Perturbations of vortices}

In order to investigate the stability of twisted vortices, 
one has to examine the dynamics of their perturbations. 
Let $\Psi(\vec{r})$ collectively denote the fields of 
a background vortex solution. 
Let us consider small perturbations around this background, 
$$
\Psi(\vec{r})\to\Psi(\vec{r})+\delta\Psi(\vec{r})\EXP{\imat\omega t}.
$$
Inserting this into the field equations \eqref{eqs} and linearizing with respect 
to the perturbations, one can put the linearized equations into a Schrodinger type 
eigenvalue problem form, 
\begin{equation}                              \label{Schr}           
(-\Delta+U)\delta\Psi=\omega^2\delta\Psi,
\end{equation} 
where the potential $U$ is determined by the background 
fields $\Psi$. 
If the spectrum of the differential operator in the left hand side of this equation  
is positive, then all the eigenfrequencies $\omega$ are real and so that the 
background
$\Psi$ 
is linearly stable. If on the other hand there are 
bound states with $\omega^2<0$, the frequency $\omega$ 
will be imaginary and the corresponding perturbation modes will grow in time as 
$e^{|\omega|t}$ thus indicating the instability of the background. 

\subsection{Special fluctuations around the ANO vortex}

Let us first briefly consider the  $q\to 0$ limit when the 
background corresponds to the embedded ANO vortex. 
It is easy to analyze a {\it particular} type of perturbations around this solution 
by simply choosing the amplitudes $f_2$ and $a_2$  in 
Eq.\eqref{ansatz1} to be small, such that the fields read  
\begin{align}\label{ansatz2}
(A_\mu+\delta A_\mu) d x^\mu&=\delta a_2(\rho)(\omega\,dt+ k\,dz)+Na_1(\rho)d\varphi,~~~~~~~~\nonumber \\
  \PHI+\delta\PHI &=\left(\begin{array}{c}f_1(\rho)\EXP{\imat N\varphi} \\
    \delta f_2(\rho)\EXP{\imat(\omega t+\eta\varphi+k z)} \end{array}\right).
\end{align}
Here we have replaced $\Omega,K,M$, respectively, par $\omega,k,\eta$ 
 in order to emphasize that 
these parameters relate to the perturbation and not to the background. 
Of course, this gives only a special type of perturbations
and not the most general one.  The perturbation equations 
are 
obtained by simply linearizing Eqs.\eqref{4ode} with respect to $f_2=\delta f_2$ 
and $a_2=\delta a_2$. The resulting linear 
equation for $\delta a_2$ decouples and one can show that its only
bounded solution is $\delta a_2=0$. The equation for $\delta f_2$ reads
\begin{align}                        \label{ANO}
%{\cal D}_2\delta f_2\equiv 
       -\frac{1}{\rho}(\rho (\delta f_2)^\prime)^\prime &+\left(\frac{(Na_1-\eta)^2}{\rho^2}
       +\beta\left(f_1^2-1\right)\right)\delta f_2
=\varepsilon \delta f_2		\,,
\end{align}
where $f_1$ and $a_1$ refer to the $N$-th background  ANO solution,
$\eta=0,1,\ldots N-1$, and 
\be
\varepsilon=\omega^2-k^2\,.
\ee
The numerical analysis reveals \cite{Hind} 
that if $\beta>1$ and for $N=1$, $\eta=0$
 there is a bound state solution of Eq.\eqref{ANO},
\be                                                 \label{bound}
\delta f_2(\rho)=\Psi_0(\rho),~~~~~~~~~\varepsilon=\varepsilon_0(\beta)<0\,,
\ee
where 
$\Psi_0(0)\neq 0$, $\Psi_0(\infty)=0$ and $\int_0^\infty |\Psi_0|^2\rho d\rho=1$. 
The eigenvalue $\varepsilon_0$ is related to the restframe value of the background 
twist $K$ in the 
$q\to 0$ limit (see Fig.\ref{Fig1}) as \cite{SL}
\be
\varepsilon_0=-K^2\,.
\ee
The frequency therefore satisfies the dispersion relation
\be                             \label{om}
\omega^2=k^2-K^2
\ee
and so the perturbations 
\be                         \label{ANOpert}
\delta\Phi_2=e^{i\omega t+i k z}\,\Psi_0(\rho)
\ee
will be growing in time if 
\be                             \label{k}
|k|<|K|.
\ee
The ANO vortices, when embedded into the semilocal theory, 
are therefore dynamically unstable \cite{Hind}. The modes \eqref{ANOpert}
depend on $z$ and so they describe inhomogeneous instabilities,
apart from the $k=0$ mode corresponding to the {\it homogeneous}
instability of the embedded vortex \cite{Hind}.  
For
\be                             \label{kk}
|k|> |K|
\ee
one has $\omega^2>0$ and Eq.\eqref{ANOpert} gives 
stationary deformations of the ANO vortex. 
%$\omega,k$ should rather be associated to the background and so  
Eq.\eqref{ansatz2} (with $\delta a_2=0$) then describes the ANO background
plus the first order correction due to the twist/current. For
$k=\pm K$ one has $\omega=0$ and so Eq.\eqref{ANOpert} gives zero modes 
corresponding to
static restframe deformations of the ANO vortex by the current.

\subsection{Generic perturbations of twisted vortices}

Since the twisted vortices are unstable in the 
 $q\to 0$ limit, when their current vanishes, one can expect that they
will presumably be unstable also for $q\neq 0$, at least for $q\ll 1$ 
when the current is small. 
However, it is not logically excluded that they may become stable for larger
values of the current.  Let us therefore consider small fluctuations around the 
generic twisted vortex configuration, 
\begin{eqnarray}                 \label{pert00}
\PHI=\PHI^{[0]}+\delta\PHI^{[0]} ,
\qquad A_\mu = A^{[0]}_\mu+\delta A^{[0]}_\mu,
\end{eqnarray}
where $\PHI^{[0]}$, $A^{[0]}_\mu$ are given by Eq.\eqref{ansatz1}. 
In order to analyze the dynamics of perturbations it 
is convenient to make use of the fact that the background configurations 
are stationary and cylindrically symmetric \cite{SL}. Since the corresponding 
symmetry generators $\partial/\partial t$, $\partial/\partial z$, 
$\partial/\partial\varphi$ 
 commute between themselves, there exists a gauge where 
the background fields do not depend on 
$t,z,\varphi$.  
 However, one  
cannot pass to this gauge using only the local U(1) gauge symmetry
that we have in our disposal, since 
adjusting {\it one} gauge function  $\Lambda(x)$ in Eq.\eqref{local} 
does not allow to eliminate the phases of the {\it two} Higgs field components at the same time. 
However, the following technical trick can be employed. 

Let us introduce 
an auxiliary SU(2) gauge field $W_\mu=\tau^a W^a_\mu$ 
 which is {\it pure gauge}. 
Replacing then in Eqs.\eqref{lagr},\eqref{eqs} the covariant derivative as    
\begin{equation}                                   \label{gauge}
D_\mu \rightarrow \mathcal{D}_\mu = D_\mu -\imat W_\mu^a\tau^a, 
\end{equation}
the 
$\mathbf{SU}(2)_{\rm global}\times \mathbf{U}(1)_{\rm local }$
symmetry \eqref{local} can be promoted to the 
$\mathbf{SU}(2)_{\rm local}\times \mathbf{U}(1)_{\rm local }$ gauge transformations: 
\begin{equation}                          \label{loc}
\PHI\rightarrow \mathbf{U}\PHI\;,\quad
A_\mu+W_\mu\rightarrow \mathbf{U}(A_\mu+W_\mu+i\partial_\mu )\mathbf{U}^{-1}, 
\end{equation}
where $\mathbf{U}=\EXP{\imat\Lambda(x)+\imat\theta_a(x)\tau^a}$. 
Since $W_\mu$ is pure gauge, one can gauge it away and then we return to the 
previous formulation of the theory. However, allowing for non-zero 
values of $W_\mu$ 
gives us an additional local SU(2) gauge freedom,
which can be useful. Although 
within the original semilocal model the field $W_\mu$ is merely 
a technical tool, and so  
we shall have to remove it at the end of  
calculations,  it can be viewed as a physical field
if 
the semilocal model is considered as the limit of the Weinberg-Salam theory. 
%then the field $W_\mu$ can be viewed as physical.  
%and so we can keep it. 

Let us now apply to Eq.\eqref{pert00}
the gauge 
transformation \eqref{loc} generated by 
\be
\mathbf{U}=\left(
\begin{array}{cc}
    e^{-\imat N\varphi}& 0 \\
   0  & e^{-\imat (\Omega t+K z+M\varphi)}
\end{array}\right). 
\ee
This transforms  
$
A^{[0]}_\mu+\delta A^{[0]}_\mu\to A^{[1]}_\mu+\delta A^{[1]}_\mu
$
and 
$
\PHI^{[0]}+\delta\PHI^{[0]}\to\PHI^{[1]}+\delta\PHI^{[1]}
$
as well as 
$
W^{[0]}_\mu=0\to W^{[1]}_\mu
$
where the new background fields 
%\be
%\PHI^{(1)}=\mathbf{U}\PHI^{(0)},~~~~
%A^{(1)}_\mu+W^{(1)}_\mu=\mathbf{U}(A^{(0)}_\mu+i\partial_\mu )\mathbf{U}^{-1}.
%\ee
%with
\begin{align}\label{ansatz3}
  \PHI^{[1]} &=\left(
\begin{array}{c}
    f_1(\rho)\\
    f_2(\rho)
\end{array}\right),~~~~
W^{[1]}_\mu d x^\mu= \frac{\tau^3}{2}(\Omega dt+K dz+(M-N)d\varphi),\notag \\
  A^{[1]}_\mu d x^\mu&=(a_2(\rho)-\frac{1}{2})(\Omega dt+ K dz)
  +(N(a_1(\rho)-\frac{1}{2})-\frac{M}{2})d\varphi,
\end{align}
while the old and new perturbations are related as 
%\be
%\delta A^{(1)}_\mu=\delta A^{(0)}_\mu,~~~~~
%\delta\PHI^{(1)}=\mathbf{U}\delta\PHI^{(0)}\,,
%\ee
%so that $\delta A_\mu$ rests invariant, while 
\be                          \label{old}
\delta A^{[1]}_\mu=\delta A^{[0]}_\mu,~~~~~
\delta\Phi_1^{[1]}=e^{-\imat N\varphi}\delta\Phi^{[0]}_1\,,~~~~~
\delta\Phi_2^{[1]}=e^{-\imat(\Omega t+Kz + M\varphi)}\delta\Phi^{[0]}_2\,.
\ee
As a result, the background fields 
depend now only on $\rho$. The price we pay for this is a new
field $W^{[1]}_\mu$ that appears in the new gauge. 
Moreover, we observe that the azimuthal components of the vector fields,
$A^{[1]}_\varphi$ and $W^{[1]}_\varphi$, do not vanish at $\rho=0$, 
which means that the 
fields are not defined at the symmetry axis.  
Nevertheless, we can work in this gauge and calculate 
$\delta A^{[1]}_\mu$, $\delta \PHI^{[1]}_\mu$,
provided that when transformed back to the old gauge
where the background fields are globally regular, the perturbations 
 $\delta A^{[0]}_\mu$ and $\delta \PHI^{[0]}_\mu$
obtained via Eq.\eqref{old} 
will also be regular.

Inserting $\PHI=\PHI^{[1]}+\delta\PHI^{[1]}$ and 
$A_\mu = A^{[1]}_\mu+\delta A^{[1]}_\mu$
to Eqs.\eqref{eqs}, replacing the covariant derivatives $D_\mu$ by 
$\mathcal{D}_\mu$,
linearizing with respect to the perturbations 
and omitting the superscript 
gives 
\begin{align}\label{lin}
&\partial_\mu\partial^\mu\delta A_\nu-\partial_\nu\partial_\mu\delta A^\mu =
%+2|\PHI|^2\delta A_\nu =  
\notag \\
&=\imat\left[(\mathcal{D}_\nu\PHI)^\dagger\delta\PHI
+(\mathcal{D}_\nu\delta\PHI)^\dagger\PHI
-\PHI^\dagger \mathcal{D}_\nu\delta\PHI 
-\delta\PHI^\dagger \mathcal{D}_\nu\PHI\right]-2|\PHI|^2\delta A_\nu\,,
\notag\\
\mathcal{D}_\mu \mathcal{D}^\mu\delta\PHI 
-2&\imat\delta A_\mu \mathcal{D}^\mu\PHI
-\imat\PHI \partial_\mu\delta A^\mu
= -\beta\left(2|\PHI|^2-1\right)\delta\PHI-\beta\delta\PHI^\dagger\PHI^2,
\end{align}
where $\mathcal{D}_\mu=\partial_\mu-i(A_\mu+W_\mu)$ and 
the background fields
$\PHI,A_\mu,W_\mu$ are given by \eqref{ansatz3}.

These equations are invariant under the U(1) gauge transformations
\be                                    \label{res}
\delta A_\mu\to \delta A_\mu + \partial_\mu \delta\Lambda(x),~~~~
\delta\PHI\to\delta\PHI+i\PHI\delta\Lambda(x),
\ee
which is the infinitesimal  version of the local U(1) 
gauge symmetry contained in \eqref{loc}.

\subsection{Separation of variables} 

Since the coefficients in Eqs.\eqref{lin} depend only on $\rho$, 
it is straightforward to separate the variables in these equations 
by making the Fourier type mode decompositions, 
\begin{align}\label{fluct}
\delta\Phi_a &=\sum_{\omega,\kappa,m} \cos(\omega t+m\varphi+
\kappa z)\left(\phi_a^{\omega,\kappa,m}(\rho)
+\imat \psi_a^{\omega,\kappa,m}(\rho)\right)
\notag\\
	&+\sin(\omega t+m\varphi+\kappa z)\left(\pi_a^{\omega,\kappa,m}(\rho)+
\imat \nu_a^{\omega,\kappa,m}(\rho)\right),\notag\\
\delta A_\mu &=\sum_{\omega,\kappa,m} \xi_\mu^{\omega,\kappa,m}(\rho)
\cos(\omega t+m\varphi+\kappa z)
+\chi_\mu^{\omega,\kappa,m}(\rho)\sin(\omega t+m\varphi+\kappa z),
\end{align}
where $a=1,2$ and $\mu=1,2,3,4$ with $x^\mu=(t,\rho,z,\varphi)$, respectively.  
One can similarly decompose the gauge function $\delta\Lambda$ in \eqref{res},
\begin{align}\label{lamb}
\delta\Lambda &=\sum_{\omega,\kappa,m} 
(\cos(\omega t+m\varphi+\kappa z)\alpha^{\omega,\kappa,m}(\rho)+
\sin(\omega t+m\varphi+\kappa z)\gamma^{\omega,\kappa,m}(\rho)),
\end{align}
and so that the gauge transformations \eqref{res} assume the form
\begin{align}                      \label{alpha}
\phi_a^{\omega,\kappa,m}&\to \phi_a^{\omega,\kappa,m},~~~~~~~~~~~~~~~~~~~~~~~~~~~~~~
\pi_a^{\omega,\kappa,m}\to \pi_a^{\omega,\kappa,m},\notag \\
\nu_a^{\omega,\kappa,m}&\to \nu_a^{\omega,\kappa,m}
+f_a\gamma^{\omega,\kappa,m},~~~~~~~~~~~
\psi_a^{\omega,\kappa,m}\to \psi_a^{\omega,\kappa,m}+f_a\alpha^{\omega,\kappa,m},\notag \\
\xi_1^{\omega,\kappa,m}&\to \xi_1^{\omega,\kappa,m}+\omega\gamma^{\omega,\kappa,m},~~~~~~~~~~~~~
\chi_1^{\omega,\kappa,m}\to \chi_1^{\omega,\kappa,m}-\omega\alpha^{\omega,\kappa,m} \notag \\
\chi_2^{\omega,\kappa,m}&\to \chi_2^{\omega,\kappa,m}+\left(\gamma^{\omega,\kappa,m}\right)^\prime,~~~~~~~~~~
\xi_2^{\omega,\kappa,m}\to \xi_2^{\omega,\kappa,m}+\left(\alpha^{\omega,\kappa,m}\right)^\prime\,, \notag \\
\xi_3^{\omega,\kappa,m}&\to \xi_3^{\omega,\kappa,m}+\kappa\gamma^{\omega,\kappa,m},~~~~~~~~~~~~~
\chi_3^{\omega,\kappa,m}\to \chi_3^{\omega,\kappa,m}-\kappa\alpha^{\omega,\kappa,m} \notag \\
\xi_4^{\omega,\kappa,m}&\to \xi_4^{\omega,\kappa,m}+m\gamma^{\omega,\kappa,m},~~~~~~~~~~~~
\chi_4^{\omega,\kappa,m}\to \chi_4^{\omega,\kappa,m}-m\alpha^{\omega,\kappa,m},
\end{align}
where 
$~~^\prime=\frac{d}{d\rho}$. 

Inserting decompositions \eqref{fluct} to the equations \eqref{lin},
the $t,z,\varphi$
variables separate and for fixed values of $\omega,\kappa,m$ one obtains
a system of 16 ODE's for the 16 real functions  
$\phi_a^{\omega,\kappa,m}(\rho)$, $\pi_a^{\omega,\kappa,m}(\rho)$,
$\nu_a^{\omega,\kappa,m}(\rho)$, 
$\psi_a^{\omega,\kappa,m}(\rho)$, $\chi_\mu^{\omega,\kappa,m}(\rho)$, 
$\xi_\mu^{\omega,\kappa,m}(\rho)$.
A further inspection  reveals 
that these 16 equations actually split into two independent
subsystems of the same size. 
Specifically, the 8 amplitudes in the left column in \eqref{alpha},
whose transformations involve only the gauge functions $\gamma^{\omega,\kappa,m}$,
satisfy a closed system of 8 equations.  Similarly, the 
amplitudes in the right column in \eqref{alpha} satisfy a closed system of 8 equations. 
These two groups of equations become identical upon the replacement
\begin{align}\label{relation1}
\pi_a^{\omega,\kappa,m}&\to\phi_a^{\omega,\kappa,m},~~~~~~
\psi_a^{\omega,\kappa,m}\to-\nu_a^{\omega,\kappa,m},~~~~~~
\xi_2^{\omega,\kappa,m}\to-\chi_2^{\omega,\kappa,m},~~~\notag\\
\chi_\mu^{\omega,\kappa,m}&\to\xi_\mu^{\omega,\kappa,m} ,~~~(\mu=1,3,4).
\end{align}
As a result, without any loss of generality we can restrict our consideration
to the 8 equations of the first group. They are explicitly listed 
in the Appendix, Eqs.\eqref{E1}--\eqref{E8}, where we have omitted 
the superscripts $\omega,\kappa,m$
to simplify the notation. As we shall see later, the two groups of equations
describe simply the real and imaginary parts of the perturbations. 

\subsection{Gauge fixing and reduction to a Schrodinger form}

One can check that Eqs.\eqref{E1}--\eqref{E8} are invariant 
under the gauge transformations expressed by \eqref{alpha}.
In addition, they satisfy the differential identity \eqref{id} 
expressing the fact that the divergence of the right hand side of
the $\delta A_\mu$ equation in Eqs.\eqref{lin} should vanish,
since the divergence of its left hand side vanishes identically. 
The identity \eqref{id} expresses one of the 8 equations in terms of 
the remaining ones,  and so only 7 equations are actually independent.
As a result, one can exclude one of the equations from consideration.

Let us now specialize to the case of purely magnetic backgrounds by 
setting 
\be
\Omega=0.
\ee
There is no loss of generality whatsoever in imposing this condition. 
Indeed, since the twisted vortices with $\Omega\neq 0$ can be obtained 
from those with $\Omega=0$ by a Lorentz boost, the same should apply
to their perturbations. It is therefore sufficient to consider  perturbations
in the $\Omega=0$ case, since  Lorentz boosting them will give
perturbations for the $\Omega\neq 0$ backgrounds. 

Next, we fix the gauge freedom \eqref{res},\eqref{alpha}  by imposing 
 the temporal gauge condition 
\be                                                                              \label{temporal} 
\delta A_0=0~~~\Leftrightarrow~~\xi_1\equiv\xi_1^{\omega,\kappa,m}=0.
\ee
This fixes the gauge completely if 
$\omega\neq 0$, although leaving a residual gauge freedom in the $\omega=0$ sector
expressed by \eqref{res} with $\delta\Lambda$ independent on time. 
Now, equation \eqref{E7} for $\xi_1$ is the Gauss constraint, and when $\xi_1=0$ it assumes
the form 
\be                                            \label{Gauss}
\omega\left(\chi_2^\prime+\frac{1}{\rho}\,\chi_2-\frac{m}{\rho^2}\,\xi_4-\kappa\xi_3
-2f_1\nu_1-2f_2\nu_2\right)=0.
\ee
We use this equation to express $\xi_3$ in terms of the other variables, 
\begin{equation}                                       \label{xi4} 
\xi_3=\frac{1}{\kappa}\left( \chi_2^\prime+\frac{\chi_2}{\rho}
-\frac{m\xi_4}{\rho^2}-2(f_1\nu_1+f_2\nu_2)\right),
\end{equation}
which is possible as long as $\kappa\neq 0$, and so that from 
now on  we can exclude $\xi_3$ from the 
consideration. Since one of the 8 equations is redundant, we can also exclude 
from the consideration 
the equation \eqref{E5} for $\xi_3$. As a result, we now have only 6 
 equations for the 6 independent field amplitudes $\phi_a$, $\nu_a$, $\chi_2$, $\xi_4$
 (see the Appendix for more details). 

Remarkably, the elimination of $\xi_3$ also truncates the residual gauge freedom
of  time independent gauge transformations. 
Specifically,  the gauge transformation \eqref{alpha} 
with $\omega=0$ leave all the equations invariant, in particular the Gauss constraint \eqref{Gauss}. 
However, the gauge invariance of the latter is only insured by the  overall factor of 
$\omega$, expressing the fact that Eq.\eqref{Gauss} is a total time derivative. 
When $\omega=0$, the whole expression   in \eqref{Gauss} is zero and so that it is gauge invariant. 
However, the expression between the parenthesis in  \eqref{Gauss}, sometimes
called `strong Gauss constraint' \cite{MV95}, is {\it not} gauge invariant under \eqref{alpha} 
for an arbitrary gauge function $\gamma\equiv \gamma^{0,\kappa,m}$ but only if 
$\gamma$ satisfies the condition
\be                                                                   \label{gamm} 
\gamma^{\prime\prime}+\frac{1}{\rho}\,\gamma^\prime=\left(\frac{m^2}{\rho^2}
+2(f_1^2+f_2^2)+\kappa^2\right)\gamma.
\ee    
Since we have used the strong Gauss constraint to express $\xi_3$, 
we have actually reduced the residual gauge freedom to the two parameter family of solutions of this 
differential equation. It is not then difficult the see that if $\gamma$ does not vanish identically, then 
it must be unbounded for small or for large $\rho$ (or in both limits). 
 The behavior of the pure gauge modes produced by this $\gamma$ 
is thus incompatible with the boundary conditions 
at the origin or at infinity (see Eq.\eqref{zero},\eqref{infty} below), unless $\gamma=0$.   
The only
admissible solution of Eq.\eqref{gamm} is therefore $\gamma=0$, which fixes the residual 
gauge freedom completely.

We finally arrive at a system of 6 independent 
second order equations. 
They are 
listed in the Appendix, Eqs.\eqref{ee1}--\eqref{ee6}, and 
they can be rewritten in the form 
\begin{align}                               \label{eig}
&-\Psi^{\prime\prime}+\mathbf{U}\Psi=\omega^2\Psi\,.
\end{align}
Here $\Psi$ is a 6-component vector 
whose components are expressed in terms of $\phi_a$, $\nu_a$, $\chi_2$, $\xi_4$
and their first derivatives, 
and 
$\mathbf{U}$ is a potential  energy matrix determined by the background solution
and depending also on $\kappa,m$. 
This determines a linear eigenvalue problem 
on a half-line, $\rho\in[0,\infty)$, the points $\rho=0,\infty$ being singular points of the 
differential equations. 

\subsection{Boundary conditions for perturbations}

Let us study the local behavior of solutions of Eqs.\eqref{eig}
for small and large $\rho$.  
It is convenient to introduce the 
 linear combinations 
\begin{align}\label{new_functions}
&\phi_a=h_a^{+}+h_a^{-},
&\nu_a=h_a^{+}-h_a^{-},\notag\\
%&\phi_2^{\omega,\kappa,m}=\frac{1}{2}\left(h_2^--h_2^+\right),
%&\nu_2^{\omega,\kappa,m}=\frac{1}{2}\left(h_2^++h_2^-\right),\notag\\
&\xi_4=\rho(g^{+}-g^{-}),
&\chi_2=g^{+}+g^{-}.
\end{align}
Using the asymptotic expansions of the background solutions at small $\rho$ \cite{SL}
to determine
the asymptotic behavior of $\mathbf{U}$ one can 
construct series solutions of Eqs.\eqref{eig} in the vicinity of $\rho=0$. 
One finds that bounded at $\rho=0$ solutions behave as 
\be\label{zero}
h_1^{\pm}=A_1^\pm\rho^{|N\mp m|}+\ldots, ~~~~~~~~
h_2^{\pm}=A_2^\pm\rho^{|M\mp m|}+\ldots, ~~~~~~~~
g^{\pm}=B^\pm\rho^{|m\mp 1|}+\ldots,
\ee
where $A_a^\pm,B^\pm$ are integration constants and the dots denote the subleading terms.  

Let us now consider the asymptotic region, $\rho\to\infty$.
Setting the 
background field amplitudes to their vacuum values, 
$f_1=a_1=1$, $f_2=a_2=0$,  so that the background fields \eqref{ansatz3} become pure gauge, 
Eqs.\eqref{ee1}--\eqref{ee6} decouple from each other and read 
(with $\D_{+}$ being defined in the Appendix)
\begin{align*}
\left(-\D_{+}+\frac{m^2}{\rho^2}+\kappa^2-\omega^2
+2\beta\right)\phi_1&=0,\\
\left(-\D_{+}+\frac{m^2}{\rho^2}+\kappa^2-\omega^2
+2\right)\nu_1&=0,\\
\left(-\D_{+}+\frac{(m\mp1)^2}{\rho^2}+\kappa^2-\omega^2
+2\right)g^\pm&=0,\\
\left(-\D_{+}
+\frac{(N-M\mp m)^2}{\rho^2}+(\kappa\pm K)^2-\omega^2\right)
h_2^\pm&=0.
\end{align*}
%where $D^2_s=-\frac{1}{\rho}\frac{d}{d\rho}\rho\frac{d}{d\rho}+\frac{m^2}{\rho^2}$
The first equation here represents a massive Higgs boson with $m_{\mbox{\tiny H}}=\sqrt{2\beta}$, 
the next three 
correspond to one longitudinal and two transverse degrees of polarization of  
a massive vector boson with $m_v=\sqrt{2}$, while the last two equations for $h_2^\pm$
describe a pair of Goldstone particles. The fluctuation equations therefore reproduce 
correctly the  physical spectrum of the semilocal model, and  
the asymptotic  behaviour of the solutions  for $\rho\to\infty$ is 
\begin{align}                           \label{infty}
\phi_1&=\frac{a_1}{\sqrt{\rho}}\,\exp\{-\mu_{\mbox{\tiny H}}\rho\}+\ldots, ~~
\nu_1=\frac{a_2}{\sqrt{\rho}}\,\exp\{-\mu_v\rho\}+\ldots, ~~\notag \\
g^\pm&=\frac{a_\pm}{\sqrt{\rho}}\,\exp\{-\mu_v\rho\}+\ldots, ~~~~~~
h_2^\pm=\frac{b^\pm}{\sqrt{\rho}}\,\exp\{-\mu_{\pm}\rho\}+\ldots\,,
\end{align}
where $a_1,a_2,a_\pm,b_\pm$ are integration constants and 
\be
\mu_{\mbox{\tiny H}}^2=\kappa^2-\omega^2+2\beta,~~~~~
\mu_v^2=\kappa^2-\omega^2+2,~~~~~
\mu^2_\pm=(\kappa\pm K)^2-\omega^2\,.
\ee

\section{Stability test}

Summarizing the above analysis, we have arrived at the eigenvalue problem \eqref{eig},
and now we wish  to find out whether it admits bound state solutions 
with $\omega^2<0$. If exist, such solutions would correspond to unstable modes of the 
background vortex configuration. 

Bound state solutions of Eqs.\eqref{eig} should be everywhere regular and so they satisfy the 
boundary conditions \eqref{zero},\eqref{infty} at the origin and at infinity. 
One possibility to proceed would then be to directly integrate Eqs.\eqref{eig} 
looking for solutions with $\omega^2<0$, which would require solving the boundary 
value problem for the 6 coupled second order differential equations.  
Fortunately, there is a faster way, 
since to know whether the negative modes  exist or not it is not
actually necessary to construct them explicitly. 
A simple method that we can apply to reveal
their existence is  to use the Jacobi criterion \cite{jack}
(see \cite{baacke} for applications of this method for the monopole stability problem). 
This method essentially relates to  
the well known fact that the ground state wave function does not oscillate, the first 
excited state has one node and so on. It follows then that if the zero energy 
solution of the Schrodinger equation oscillates, then the ground state energy eigenvalue
is negative. When applied to multichannel systems as in our case,  the Jacobi method
gives the following recipe. 

Let $\Psi_{(s)}(\rho)$ where 
$s=1,\ldots 6$ be 6 linearly independent and regular at the origin 
solutions of Eqs.\eqref{eig} with $\omega^2=0$, each 
of them being a 6-component vector $\Psi^I_{(s)}(\rho)$.
These solutions can be chosen to satisfy, for example, the conditions 
$\Psi^I_{(s)}(\rho_0)=\delta^I_s$,
where $\rho_0$ is a point close to the origin. 
Equally, each of them can be obtained by 
taking the boundary conditions \eqref{zero}, setting to zero 5 out of the 6 integration 
constants in \eqref{zero}, and integrating numerically Eqs.\eqref{eig} with 
$\omega^2=0$ 
towards large values of $\rho$. 
Now, if the determinant 
\begin{equation}                            \label{det} 
\Delta(\rho)=\left|\begin{array}{ccc}
\Psi^1_{(1)}(\rho) &\dots&\Psi_{(6)}^1(\rho) \\
\vdots&\ddots&\vdots\\
\Psi_{(1)}^6(\rho) &\dots&\Psi_{(6)}^6(\rho) 
\end{array}\right|
\end{equation}
vanishes somewhere,  then there exists a negative part of the spectrum. 
Thus we can decide whether the backgrounds is stable or not by simply calculating 
the determinant \eqref{det} -- a much easier task than   
solving  the boundary value problem \eqref{eig} to directly find the eigenvalues.  
According to \cite{Amm}, the number of instabilities is equal 
to the number of nodes  of $\Delta(\rho)$.

\subsection{Jacobi test in the  ANO limit}

Before studying the general case, let us consider the limit where $q=f_2=a_2=0$ 
and the twisted vortices reduce to the embedded ANO solutions. We already know that 
in this case there is a particular perturbation 
\eqref{ansatz2} leading  to the eigenvalue problem \eqref{ANO}
admitting the negative mode \eqref{ANOpert}. 
Now we shall be able to 
take into account all the remaining perturbation modes. 
The perturbation equations in the ANO limit split into  the  system of four coupled equations 
\eqref{NO} plus two decoupled equations  \eqref{SS}.

\begin{figure}[ht]
\hbox to\linewidth{\hss%
	\psfrag{rho}{$\rho$}
	\psfrag{D}{}
	\psfrag{Dn1m0}{$\Delta_{1,0}$}
	\psfrag{Dn1m1}{$\Delta_{1,1}$}
	\psfrag{Dn1m2}{$\Delta_{1,2}$}
	\psfrag{Dn2m2}{$\Delta_{2,2}$}
	\psfrag{hplus}{$h_2^{+}$}
	\resizebox{8cm}{6cm}{\includegraphics{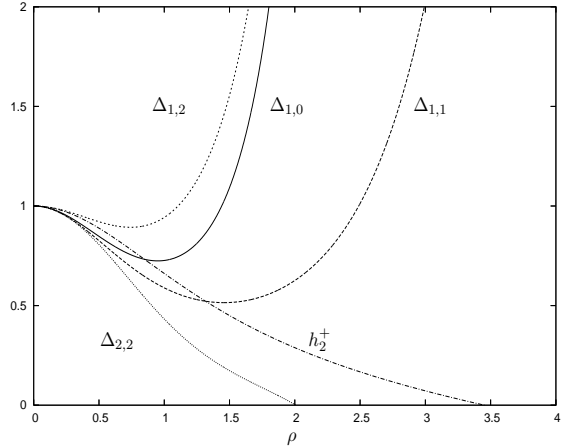}}
\hss}
\caption{Jacobi determinant $\Delta_{N,m}$
for solutions of Eqs.\eqref{NO} with $\omega=\kappa=0$
and $h_2^{+}$ given by \eqref{SS} with $\omega=m=0$, $\kappa=K$
in the case of the $\beta=2$ ANO background with $N=1$. 
Since $\Delta_{1,m}$ do not vanish, the $N=1$ ANO vortex is stable
within the original ANO theory. The vanishing of $\Delta_{2,2}$ indicates the splitting 
instability of the $N=2$ vortex. The vanishing of  $h_2^{+}$ shows  the  instability
of the $N=1$ vortex embedded into the semilocal theory. 
}
\label{Fig2}
\end{figure}
 
The four equations \eqref{NO}  
do not contain perturbations of the second component of the Higgs field
and so they actually describe the dynamics of the perturbed ANO vortex
within the original one component ANO model. We therefore expect that for $N=1$ these 
equations do not admit bound state solutions with $\omega^2<0$, since 
the ANO vortex is stable within the original ANO theory. However, since for $\beta>1$ 
the multivortex configurations  of higher winding numbers are unstable 
with respect to splitting into vortices of lower winding numbers \cite{BV},\cite{Good}, 
 Eqs.\eqref{NO} should admit 
bound state solutions with $\omega^2<0$ for $\beta>1$ and  for $N\geq 2$. 
We can now test these assertions using the Jacobi criterion. 
Integrating Eqs.\eqref{NO} with $\omega^2=0$ with the regular at the origin 
boundary conditions \eqref{zero} to construct their four linearly independent solutions,
we calculate the determinant $\Delta(\rho)$ as in Eq.\eqref{det}. 
It turns out then that  $\Delta(\rho)$ is indeed always positive for $N=1$, while for
 $N=2$ and $\beta>1$ it changes sign for quadrupole perturbation modes with $m=2$  
(see Fig.\ref{Fig2}). All this agrees with the expected results.

Let us now consider the two decoupled equations \eqref{SS}. 
They have exactly the same structure as Eq.\eqref{ANO}, up to the replacement 
\be
\delta f_2\to h_2^{\pm},~~~~~
\eta\to M\pm m,~~~~~
k\to K\pm\kappa\,.
\ee
We therefore immediately know that for $N=1$, in which case $M=0$, choosing $m=0$,
these equations admit bound state solutions. 
We can also directly apply the Jacobi criterion to Eqs.\eqref{SS}. Setting $\omega=0$
and also $K+\kappa=0$ (or $K-\kappa=0$) and integrating gives a solution for $h_2^{+}$ 
(or for $h_2^{-}$) 
that passes through zero once
(see Fig.\ref{Fig2}) thus showing that there is exactly one bound state. 

\subsection{The $N=1$ twisted vortices}

\begin{figure}[ht]
\hbox to\linewidth{\hss%
\psfrag{rho}{$\rho$}
	\psfrag{D}{}
	\psfrag{q0}{$q=0$}
	\psfrag{q035}{$\Delta_{1,0}$}
	\resizebox{8cm}{6cm}{\includegraphics{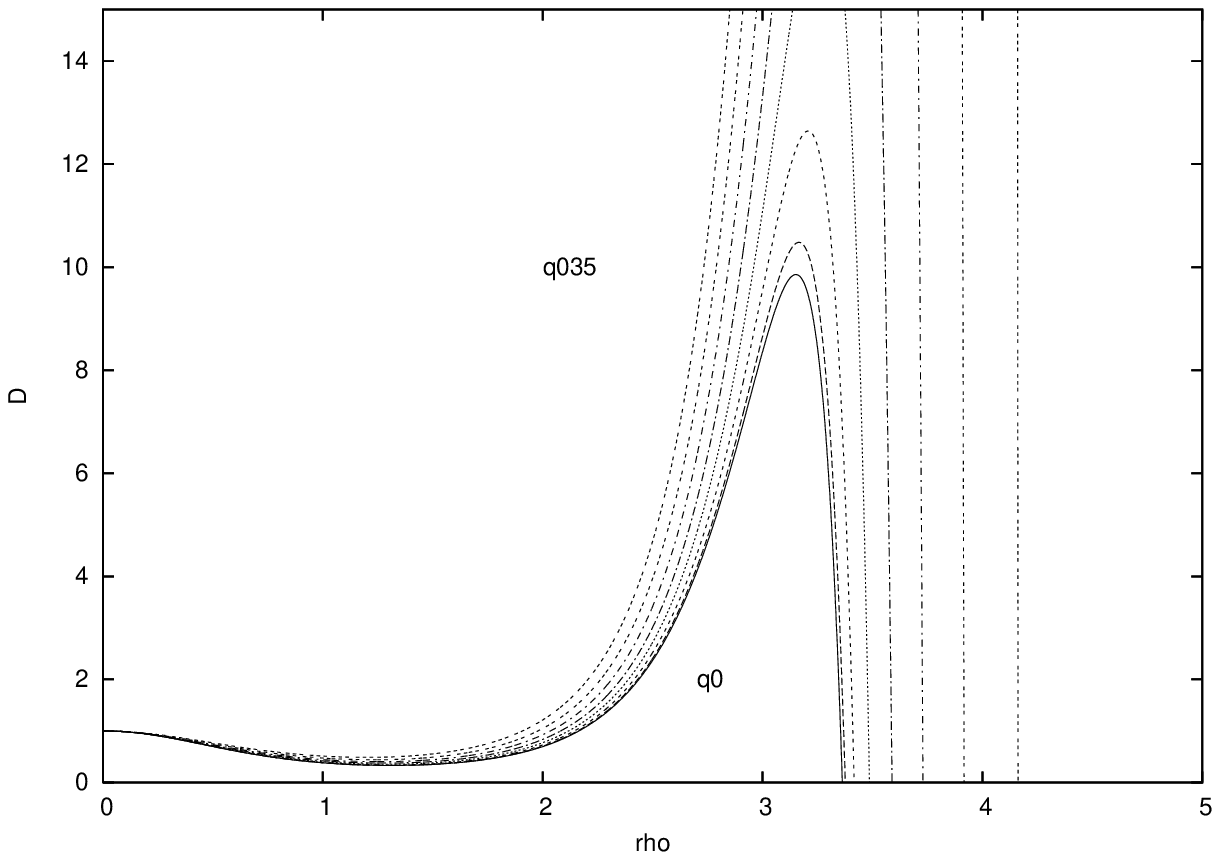}}
	
\hspace{5mm}%

\psfrag{rho}{$\rho$}
	\psfrag{D}{}
	\psfrag{Dn1m1}{$\Delta_{1,1}$}
	\psfrag{Dn1m2}{$\Delta_{1,2}$}
	\resizebox{8cm}{6cm}{\includegraphics{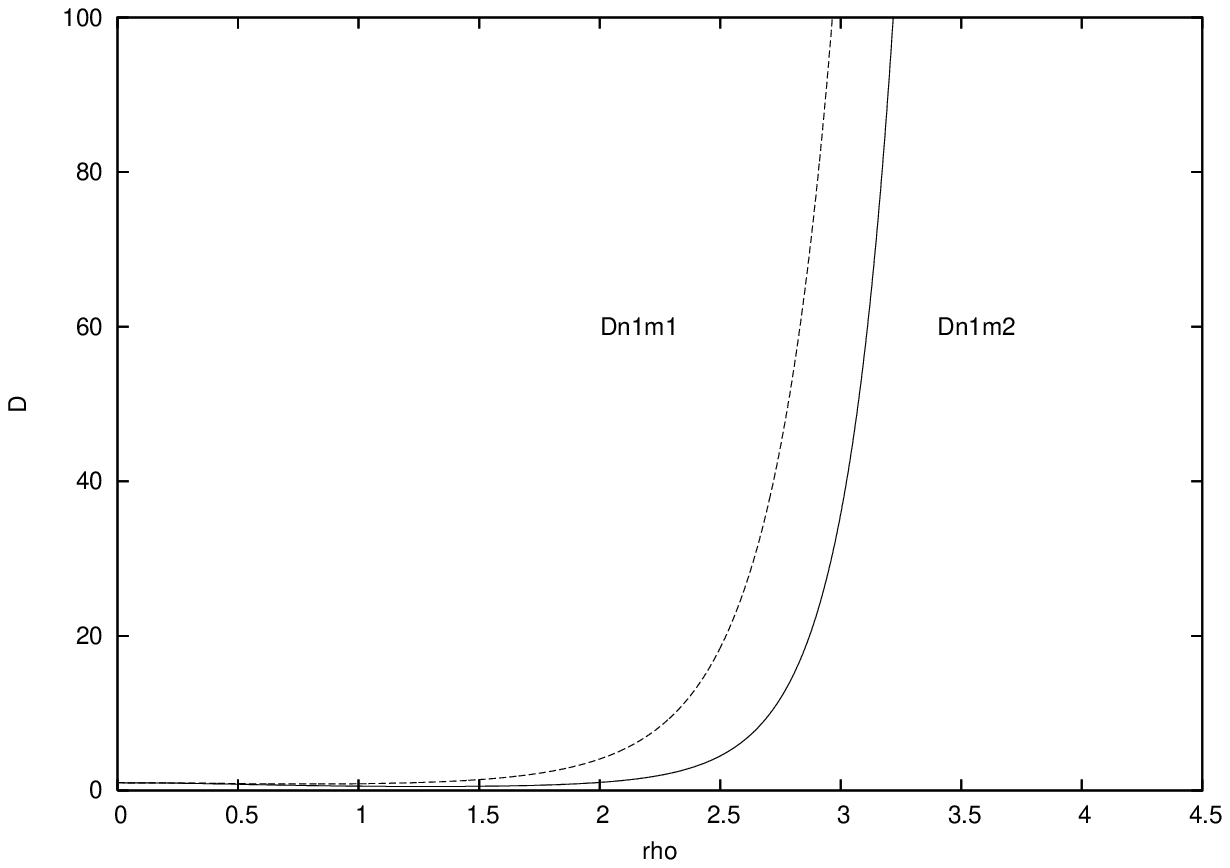}}

\hss}
	%\resizebox{10cm}{6.5cm}{\includegraphics{jacobi_N=1_m=1_m=2.ps}}
	\caption{
Right: 
$\Delta_{1,0}$ for 
$\kappa=K$ for the twisted backgrounds with $\beta=2$ and 
$q\in \left[0;0.35\right]$. The vanishing of $\Delta_{1,0}$ indicates 
the existence of a negative mode. 
Left: perturbations of the $N=1$ twisted vortex in the $m>0$ sector.
	No instabilities appear.
	Here $\beta = 2$, $q=0.5$, $\kappa = K$.  
}
\label{Fig3}
\end{figure}

We now have all the necessary ingredients to check the stability of generic
twisted vortices. To this end we integrate numerically the system of 6 coupled
equations \eqref{ee1}--\eqref{ee6} with $\omega^2=0$ with the regular at the origin
boundary conditions \eqref{zero} to construct the Jacobi determinant. 
We consider first of all the case of the fundamental $N=1$ twisted vortex,
and for the perturbation parameters we choose  the same values which give
rise to the instability in the ANO limit: $m=0$ and $\kappa=K$. 
The resulting Jacobi determinant $\Delta(\rho)$ for several values of the 
background condensate parameter $q$ is shown in Fig.\ref{Fig3}. We find that not only 
for $q\to 0$ (the ANO limit) but also for all other values 
in the interval $0\leq q<1$ the determinant crosses zero. 
All these solutions are therefore unstable with respect to axially symmetric perturbations,
and we have gone up to $\beta=10$ to check that  the number of instabilities,
given by number of zeroes of   $\Delta(\rho)$, is exactly one. 

We have also checked the sectors with $m>0$ but found no instabilities there.
In Fig.\ref{Fig3} the Jacobi determinants for $m=1,2$ are shown, they are everywhere positive. 
This is  not surprising, since increasing $m$ increases the centrifugal energy
thus rendering less probable the existence of bound states. 
Summarizing, all instabilities of the fundamental $N=1$ twisted vortex reside in the axially symmetric
$m=0$ sector.

\subsection{Solutions with $N>1$}

\begin{figure}[ht]
\hbox to\linewidth{\hss%
\psfrag{rho}{$\rho$}
	\psfrag{D}{}
	\psfrag{q0}{$q=0$}
	\psfrag{q035}{$\Delta_{2,0}$}
	\psfrag{Dn2m1}{$\Delta_{2,1}$}
	\psfrag{Dn2m3}{$\Delta_{2,3}$}
	\resizebox{8cm}{6cm}{\includegraphics{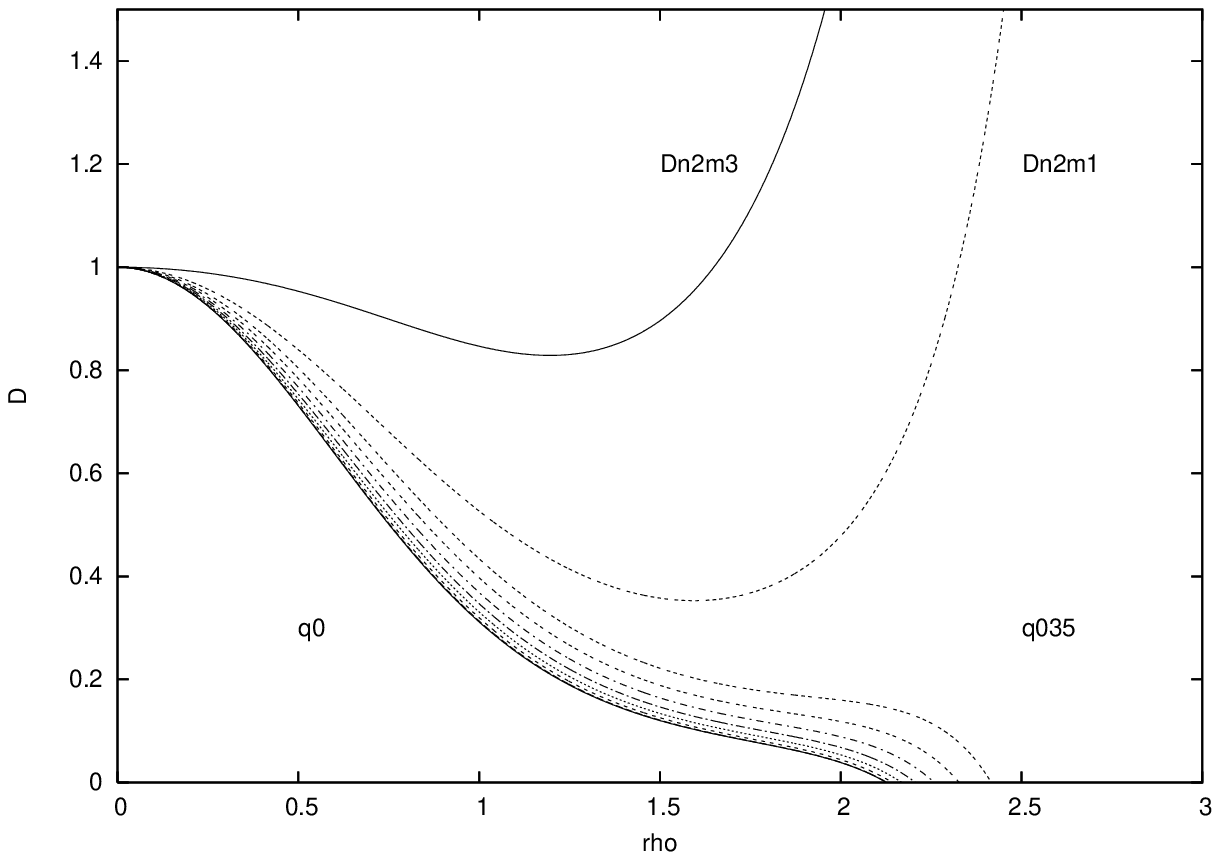}}
\hspace{5mm}%
\psfrag{rho}{$\rho$}
	\psfrag{D}{}
	\psfrag{q0}{$q=0$}
	\psfrag{q035}{$\Delta_{2,2}$}
	\resizebox{8cm}{6cm}{\includegraphics{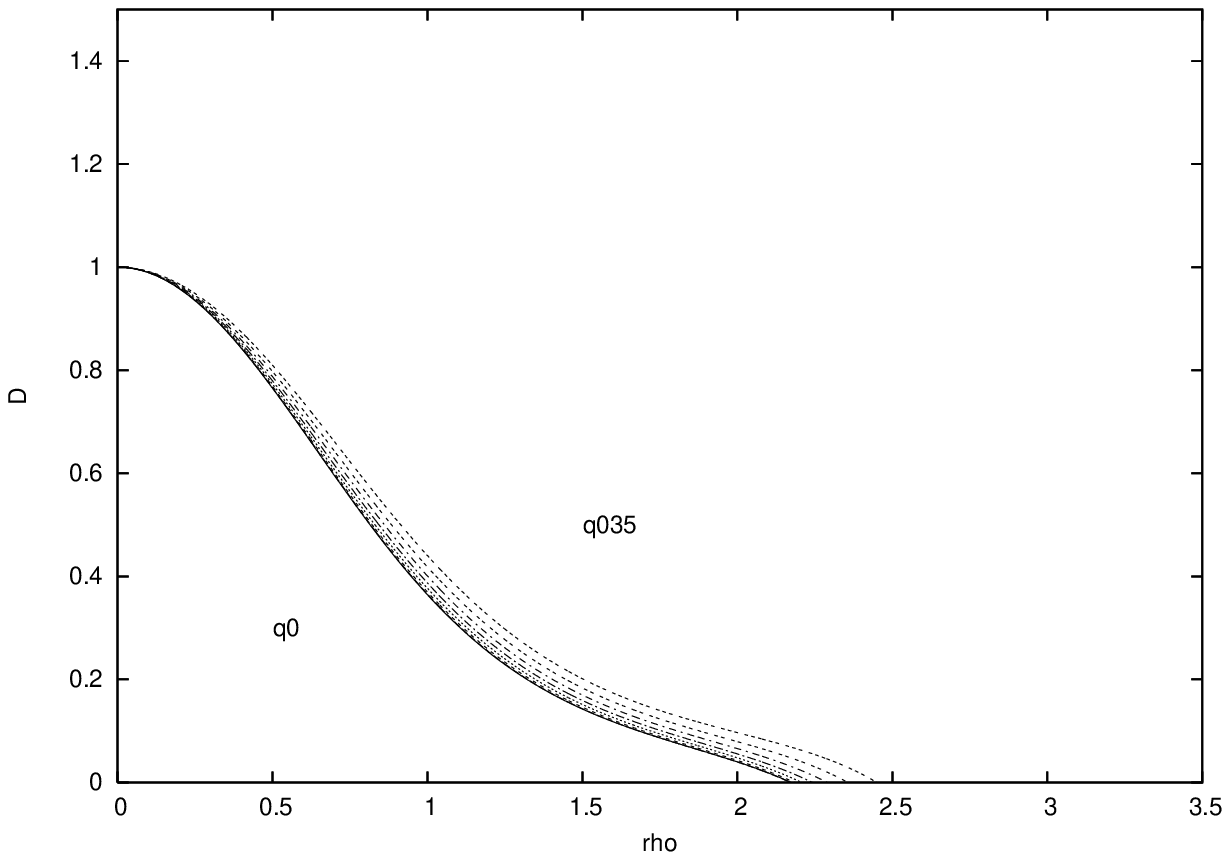}}	
\hss}
	\caption{
Left: perturbations of the $N=2$ vortices. Here
$\beta = 2$, $\kappa = K$ and $q\in \left[0;0.35\right]$. The same $m=0$
instability as for the $N=1$ vortex is detected. 
$N=2$ vortices  are however stable with respect to perturbations
	with $m>0$, $m\neq 2$. Here 
$\beta = 2$, $\kappa = K$ and $q=0.5$. 
Right: they are unstable with respect to the $m=2$ quadrupole deformations -- 
splitting instability. Here 
$\beta = 2$, $\kappa = K/2$ and $q\in \left[0;0.35\right]$.}
\label{Fig4}
\end{figure}

Vortices with $N>1$ 
have additional instabilities. First, they 
have the same instability in the $m=0$ sector
as the $N=1$ solutions. This can be checked by calculating the Jacobi determinant 
for the $m=0$ perturbations; see Fig.\ref{Fig4}. In addition, solutions with $N>1$ 
can be unstable with respect to splitting  
to vortices of lower winding number. 

The latter instability is present already in the ANO limit (see Fig.\ref{Fig2}).  
In the ANO case the existence of this  splitting instability can also be inferred 
from the energy considerations:  
in the simplest case is suffices to compare the energy of the $N=2$ vortex to the 
doubled energy of the $N=1$ vortex. It turns out then that the former is higher than the latter
for $\beta>1$, which means that the vortex splitting is energetically favoured. 

A similar energy argument can also be applied to the twisted vortices.  
However, the consideration in this case is complicated by the fact that, 
apart from the winding number $N$, the twisted vortices also carry additional parameters,
in particular the current ${\cal I}$.  One can also 
compare the energy of a $N=2$ twisted vortex, 
$E(N=2,{\cal I})$, with a sum of energies of two $N=1$ vortices,
$E(N=1,{\cal I}_1)+E(N=1,{\cal I}_2)$, but then one should 
consider different possibilities for
 values of ${\cal I}_1$ and ${\cal I}_2$.  Assuming that 
${\cal I}_1+{\cal I}_2={\cal I}$ and choosing 
${\cal I}_1={\cal I}_2={\cal I}/2$ one finds that the splitting is indeed energetically favoured 
if ${\cal I}$ is small, but not for large enough ${\cal I}$ \cite{SL}. 
However, one can also consider the case where ${\cal I}_1$ is large and positive, while 
${\cal I}_2$ is large and negative, their sum ${\cal I}_1+{\cal I}_2={\cal I}$ being fixed. 
Since the energy of twisted vortices decreases with the current (see Fig.\ref{Fig1}), 
it follows that whatever the value of 
${\cal I}$ is, one can always adjust ${\cal I}_1$ and ${\cal I}_2$ such that the splitting will be 
energetically favorable. 

We therefore expect all twisted vortices with $N>1$ to be unstable with 
respect to the splitting.
This expectation is confirmed by calculating the  Jacobi determinant  
in the $m=2$ sector, since 
for all $N=2$ backgrounds we tested  
the determinant vanishes 
somewhere; see Fig.\ref{Fig4}. 
We have also checked that $N=3$ vortices exhibit the splitting 
instability for $m=2,3$. 

Summarizing what has been said, $N>1$ current carrying vortices 
exhibit the same
axially symmetric  $m=0$ instability
as the fundamental $N=1$ vortex. 
In addition they are unstable with respect to splitting into
vortices of lower winding number. 
We have not found negative modes in sectors with $m=1$  or for $m>N$. 

\subsection{Explicit computation of the eigenvalue}

Having revealed the existence of the negative modes in the spectrum of the 
fluctuation operator, we have also managed to construct them explicitly. 
Such a construction is considerably more involved than applying the Jacobi criterion, 
since it requires solving the boundary value problem for the 6 coupled equations 
\eqref{ee1}--\eqref{ee6} with the boundary conditions \eqref{zero} and \eqref{infty}.
In addition, one should solve at the same time the 4 equations 
\eqref{4ode} to generate 
the background profiles.

\begin{figure}[ht]
\hbox to\linewidth{\hss%
\psfrag{kappa}{$\kappa$}
	\psfrag{omega}{}
	\psfrag{om1no}{$\omega^2_{(1)}~~~~~~~~~~~~~~~~~~~~~~~q=0$}
	\psfrag{om2no}{$\omega^2_{(2)}$}
	\psfrag{omq1}{}
	\psfrag{omq2}{$q=0.15$}
	\psfrag{omq3}{}
	\psfrag{omq4}{$q=0.45$}
	\psfrag{omq5}{}
	\psfrag{omq6}{$q=0.55$}
	\resizebox{8cm}{6cm}{\includegraphics{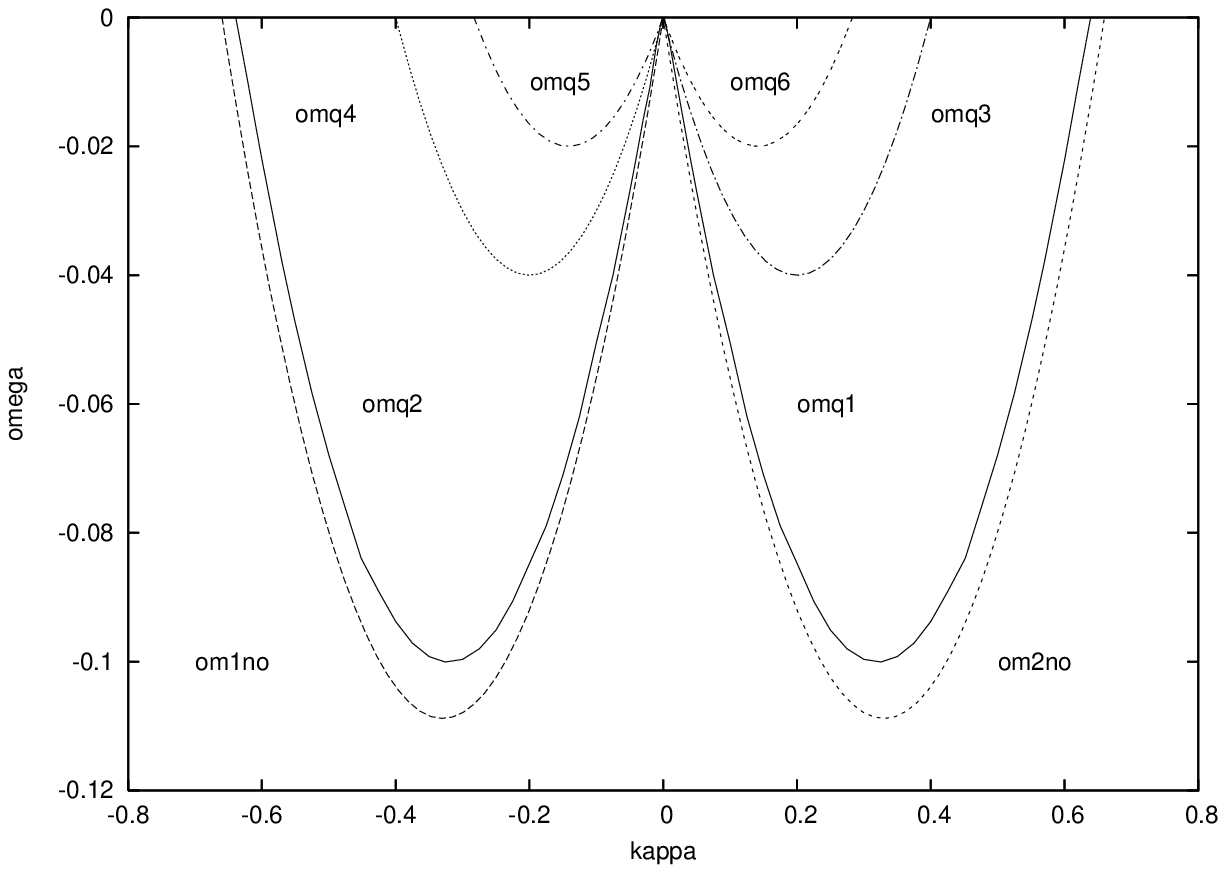}}
\hspace{5mm}%
	\psfrag{q}{$q$}
	\psfrag{omega}{}
	\psfrag{omega2}{$\omega^2$}
	\psfrag{K2}{$-K^2$}
	\resizebox{8cm}{6cm}{\includegraphics{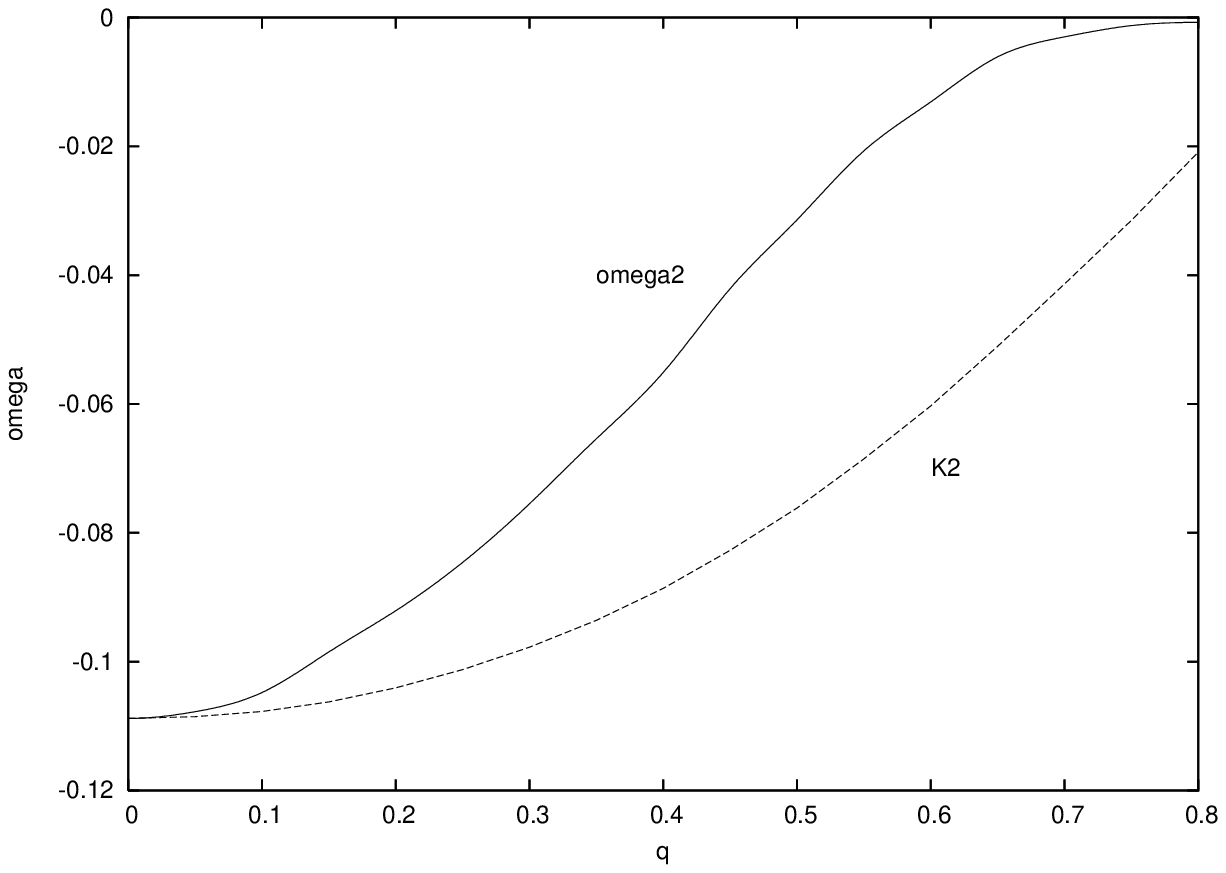}}
\hss}	
	%\resizebox{10cm}{6.5cm}{\includegraphics{reldisp.ps}}
	\caption{Left: dispersion relations $\omega^2(\kappa)$ for the two bound state solutions
of Eqs.\eqref{ee1}--\eqref{ee6} for the $N=1$ twisted backgrounds with $\beta=2$. 
Right: The minimal value of $\omega^2(\kappa)$ corresponding to $|\kappa|=K$.
We see that the instability region bounded from below by the $\omega^2(\kappa)$ curve 
shrinks when $q$ increases. 
}
\label{Fig5}
\end{figure}

It turns out that for the $N=1$ background  
Eqs.\eqref{ee1}--\eqref{ee6}  admit two different bound state solutions in the $m=0$ sector
with eigenvalues $\omega_{(1)}^2$ and  $\omega_{(2)}^2$, where  
$\omega_{(1)}^2(\kappa)<0$ for  $\kappa\in(-2K,0)$ and 
$\omega_{(2)}^2(\kappa)<0$ for  $\kappa\in(0,2K)$, also 
$\omega_{(1)}^2(\kappa)=\omega_{(2)}^2(-\kappa)$. 
As a result, for each value of $\kappa\in (-2K,0)\cup (0,2K)$ there is only one 
negative mode. For these two bound states all the 6 field amplitudes in 
Eqs.\eqref{ee1}--\eqref{ee6} are coupled, but in the limit $q\to 0$ only 2 non-trivial amplitudes
remain and the solutions reduce to that for the two decoupled equations \eqref{SS}.

Fig.\ref{Fig5} 
shows the dispersion relations $\omega^2_{(1)}(\kappa)$ and $\omega^2_{(2)}(\kappa)$
for several values of $q$. For $q=0$ they are given by the formula
\be                                                 \label{ome}
\omega^2=(\kappa\pm K)^2-K^2
\ee
which will be explained shortly. For $q>0$ this formula is no longer exact, but it approximates
reasonably well the values of $\omega^2(\kappa)$. (This formula was used to trace the 
$q=0.45$ and $q=0.55$ curves in Fig.\ref{Fig5}, since the direct calculation of 
$\omega^2(\kappa)$, performed for $q=0.15$, becomes rather time consuming 
for larger values of $q$.)
The following features of this formula remain true also for $q>0$. 
The negative mode exists only for $\kappa\in (-2K,0)\cup (0,2K)$. 
The minimal value of 
$\omega^2$ is achieved for $\kappa=\pm K$ and this value tends to zero as $q\to 0$,
being bounded from below by $-K^2$ (see Fig.\ref{Fig5}). 
We observe that the  
size of the instability region rapidly shrinks  for $q\to 1$.  
 The twisted vortices become therefore `more stable'
for large currents. This can be understood by noting that their energy 
decreases with growing ${\cal I}$ and for ${\cal I}\to\infty$ 
approaches the  lower bound $2\pi |N|$ \cite{SL} and  so that less room for instability is left 
for large  ${\cal I}$. 

 \begin{table}[ht]
\begin{center}
\parbox{15cm}{\caption{The value of $\omega^2$ for $\kappa=K$, $N=1$.}}\vskip 0.3cm
\begin{tabular}{c||cccccccccccccc}
\hline
$q$ & 0 & 0.05 & 0.1 & 0.15 & 0.2 & 0.25 & 0.3 \\
$K$ & 0.3298 & 0.3294 & 0.3282 & 0.3259 & 0.3225 & 0.3181 & 0.3126 \\
$\omega^2$ & -0.108 & -0.107 & -0.105 & -0.099 & -0.092 & -0.085 & 0.076 \\
\hline
$q$ & 0.35 & 0.4 & 0.45 & 0.5 & 0.55 & 0.6 & 0.65 \\
$K$ & 0.3059 & 0.2976 & 0.2876 & 0.2759& 0.2617 & 0.255 & 0.2259 \\
$\omega^2$ & -0.065  & -0.055 & -0.042& -0.031 & -0.0207 & -0.013 & -0.007 \\
\hline
\end{tabular}
\end{center}
\label{Tab}
\end{table}

\section{Physical manifestations of instability}

Summarizing the above analysis, despite their lower energy, 
the twisted vortices have essentially the same instabilities as
the embedded ANO vortices. However, there is one important difference leading to 
important consequences: they do not have the uniform 
spreading instability mode independent of $z$
analogues to the one obtained by setting $k=0$ in Eq.\eqref{ANOpert}.
To understand how this comes about and why this is so important, let us explicitly 
reconstruct the fields corresponding to the unstable modes 
of the fundamental $N=1$ background vortex.

\subsection{Reconstructing the perturbations} 

 Let $\Psi_{(s)}$ with $s=1,2$
denote the two linearly independent {\it normalized} 
bound state solutions of the eigenvalue problem $\eqref{eig}$
for a given $\kappa$ and with $m=0$, the corresponding 
eigenvalues being $\omega_{(s)}^2$. 
Each of them determines 
the 6 field amplitudes $\phi_a^{(s)}$, $\nu_a^{(s)}$, $\xi_4^{(s)}$, $\chi_2^{(s)}$
in Eqs.\eqref{ee1}--\eqref{ee6}
and these, using \eqref{xi4}, determine also $\xi_3^{(s)}$. 
Let us  introduce vectors 
\be
X^{\pm}_{(s)}=\left(\begin{array}{c}
\phi_a^{(s)} \\
\pm\nu_a^{(s)} \\
\xi_\mu^{(s)} \\
\pm\chi_2^{(s)}
\end{array}\right),~~~~~
Y_{\omega,\kappa}=\left(\begin{array}{c}
\phi_a^{\omega,\kappa} \\
\nu_a^{\omega,\kappa} \\
\xi_\mu^{\omega,\kappa} \\
\chi_2^{\omega,\kappa}
\end{array}\right),~~~~~
Z_{\omega,\kappa}=\left(\begin{array}{c}
\pi_a^{\omega,\kappa} \\
\psi_a^{\omega,\kappa} \\
\chi_\mu^{\omega,\kappa} \\
\xi_2^{\omega,\kappa}
\end{array}\right),~
\ee
where $\mu=3,4$ and where the components of the vectors $Y_{\omega,\kappa}$ and 
$Z_{\omega,\kappa}$
are the $m=0$ amplitudes entering the mode decomposition \eqref{fluct}.
If  $Y_{\omega,\kappa}$ describes a bound state solution of the perturbation equations,
then it should be a linear combination of $X^{+}_{(s)}$, since the latter can be regarded as
 the basis vectors in the space of bound state solutions. Similarly, if $Z_{\omega,\kappa}$  
describes a bound state solution of the second group of perturbation equations, then 
it should be a linear combination of the corresponding basis vectors. The latter, 
in view of the symmetry \eqref{relation1}, can be chosen to be $X^{-}_{(s)}$.
As a result, 
\be
Y_{\omega,\kappa}=\sum_s a^\kappa_{(s)} X^{+}_{(s)},~~~~~~~~
Z_{\omega,\kappa}=\sum_s b^\kappa_{(s)} X^{-}_{(s)},
\ee
where $a^\kappa_{(s)}$ and $b^\kappa_{(s)}$ are independent real coefficients. 
Since the value of $\omega^2$ is fixed for the bound states, the frequency 
assumes only two values, $\omega=\pm\sqrt{\omega^2}$, and one has similarly 
\be
Y_{-\omega,\kappa}=\sum_s c^\kappa_{(s)} X^{+}_{(s)},~~~~~~~~
Z_{-\omega,\kappa}=\sum_s d^\kappa_{(s)} X^{-}_{(s)}.
\ee
Inserting this to \eqref{fluct} and applying \eqref{old} 
gives the perturbations in the regular gauge, 
\begin{align}                       \label{pert}
\delta\Phi_1^{[0]}&=e^{iN\varphi}\sum_{\kappa}\sum_s
\left(h^{(s)+}_{1}
{\cal Q}^\kappa_{(s)}+
h^{(s)-}_{1}({\cal Q}^\kappa_{(s)})^\ast\right),~~\notag \\
\delta\Phi_2^{[0]}&=e^{iM\varphi+iKz}\sum_{\kappa}\sum_s\left(h^{(s)+}_{2}
{\cal Q}^\kappa_{(s)}+
h^{(s)-}_{2}({\cal Q}^\kappa_{(s)})^\ast\right),~~\notag \\
\delta A_2^{[0]}&=\sum_{\kappa}\sum_s i\left(
({\cal Q}^\kappa_{(s)})^\ast-{\cal Q}^\kappa_{(s)}\right)\chi^{(s)}_2, \notag \\
\delta A_\mu^{[0]}&=\sum_{\kappa}\sum_s\left
({\cal Q}^\kappa_{(s)}+({\cal Q}^\kappa_{(s)})^\ast\right)\xi^{(s)}_\mu,~~~~~~~\mu=3,4.  
\end{align}
Here $h^{(s)\pm}_{a}=\frac12(\phi_a^{(s)}\pm\nu_a^{(s)})$ and 
\be
{\cal Q}^\kappa_{(s)}=(A^\kappa_{(s)} \exp\{i\omega_{(s)} t\}+
B^\kappa_{(s)} \exp\{-i\omega_{(s)} t\}) \exp\{i\kappa z\}
\ee
with $A^\kappa_s =a^\kappa_s -ib^\kappa_s$ and  
$B^\kappa_s =c^\kappa_s +id^\kappa_s$.
Regularity of the perturbations \eqref{pert}
at $\rho=0$ is guaranteed by the boundary conditions
\eqref{zero}. 

\subsection{Instabilities of the embedded ANO vortex}
 
Let us consider again the ANO limit. We know that the bound states are then described by solutions
of  the two decoupled equations \eqref{SS}, 
while Eqs.\eqref{NO} do not have 
bound state solutions for $N=1$. As a result, 
one has $h^{(s)\pm}_{1}=\chi^{(s)}_2=\xi^{(s)}_\mu=0$
and, for $s=1$, 
\be                        \label{s1}
h^{(1)+}_{2}=\Psi_0(\rho),~~~~h^{(1)-}_{2}=0,~~~~~~\omega_{(1)}^2=k_{+}^2-K^2, 
\ee
while for $s=2$ 
\be                            \label{s2}
h^{(2)+}_{2}=0,~~~~h^{(2)-}_{2}=\Psi_0(\rho),~~~~~~\omega_{(2)}^2=k_{-}^2-K^2\,,
\ee
where 
$\Psi_0(\rho)$ is the same as in Eq.\eqref{bound} and $k_{\pm}=\kappa\pm K$. 
This explains, in particular, the formula \eqref{ome}. 
The only non-zero perturbation in
Eqs.\eqref{pert} is then 
\begin{align}
\delta\Phi_2&=e^{iKz}\Psi_0(\rho)\sum_{\kappa}\left({\cal Q}_{\kappa(1)}+
{\cal Q}^\ast_{\kappa(2)}\right)
\end{align}
or explicitly 
\begin{align}
\delta\Phi_2&=\Psi_0(\rho)\sum_{\kappa}
\left(\left[
A^\kappa_{(1)}\exp\left\{i\sqrt{\omega_{(1)}^2}\,t\right\}+
B^\kappa_{(1)}\exp\left\{-i\sqrt{\omega_{(1)}^2}\,t\right\} \right]
\right.
\exp\{ik_{+}z\}                      \notag \\
&+\left.\left[
(A^\kappa_{(2)})^\ast\exp\left\{-i\sqrt{\omega_{(2)}^2}\,t\right\}+
(B^\kappa_{(2)})^\ast\exp\left\{i\sqrt{\omega_{(2)}^2}\,t\right\}\right]
\exp\{-ik_{-}z\}\right).                                  \label{pert1}
\end{align}
This expression 
contains exponentially growing in time terms if
$k_\pm^2-K^2=
\kappa(\kappa\pm 2K)<0$, which condition is satisfied for 
$\kappa\in (-2K,0)\cup (0,2K)$. 
Since these terms are proportional to $\exp(\pm ik_\pm z)$, 
the instability can be viewed as a superposition of standing waves 
of wavelength $\lambda=2\pi/k_\pm$ 
whose amplitude grows in time. The minimal wavelength is 
$
\lambda_{\rm \min}={2\pi}/{K}
$
and this suggests that the instability can be removed by imposing 
the periodic boundary conditions along $z$-axis with the period 
$L<\lambda_{\rm \min}$. However, this {will not remove} 
the  particular unstable modes with 
$k_{\pm}=0$ independent on $z$ since it can be considered as
periodic with any period. Setting in \eqref{pert1} $k_{+}=0$ or $k_{-}=0$ 
shows that these modes are proportional to each other and so in fact 
there is only one such homogeneous mode,
\be                   \label{homogen}
\delta\Phi_2={A}\,\Psi_0(\rho)\exp(Kt),
\ee
with ${A}\in\mathbb{C}$. This mode describes a uniform spreading 
of the vortex in the $x,y$ plane.

Terms with $|\kappa|>2K$ in \eqref{pert1} describe 
waves travelling towards positive and negative values of $z$ with positive or 
negative frequency.  These modes correspond to stationary
deformations of the embedded ANO vortex by the twist.
In particular, setting  $|k_\pm|=K$ and so $\kappa=0,\pm 2K$
gives zero modes corresponding
to static deformations of the vortex by the twist/current,
\be                                         \label{def}
\delta\Phi_2=\Psi_0(\rho) (A_1\,e^{iKz}+A_2\, e^{-iKz}),
\ee
two independent modes here corresponding to two possible directions of the current. 
%This formula can also be obtained by differentiating the background fields \eqref{ansatz1} 
%with respect to the 

It is worth noting that the elementary waves in \eqref{pert1}
decouple one from another. In particular, denoting $k=k_{+}$ one can 
adjust the parameters $A^\kappa_{(s)}$ and $B^\kappa_{(s)}$ such that the result
will agree with Eq.\eqref{ANOpert}.

\subsection{Generic case}

Let us now consider perturbations of generic twisted
vortices given by \eqref{pert}. As in the ANO limit, there are 
two linearly independent bound state solutions of 
Eqs.\eqref{ee1}--\eqref{ee6} with eigenvalues $\omega_{(1)}^2(\kappa)$
and $\omega_{(2)}^2(\kappa)$, where $\omega_{(1)}^2(\kappa)<0$
for $-2K<\kappa<0$ and $\omega_{(2)}^2(\kappa)<0$
if $0<\kappa<2K$ (see Fig.\ref{Fig5}).  
The perturbations are therefore given by Eqs.\eqref{pert} where
the sums contain growing in time terms for 
$\kappa\in (-2K,0)\cup (0,2K)$. 
(When $\kappa\to0,\pm 2K$ the solutions approach zero modes analogues to \eqref{def}
and describing static on-shell  deformations inside the family of twisted vortex
solutions.  In the generic case such zero modes contain a term linear in $z$, 
 as can be seen by differentiating the background fields \eqref{ansatz1} with respect to $K$,
and so the solutions approach these modes only pointwise and non-uniformly in $z$.)

The novel feature as 
compared to the ANO case is that now Eqs.\eqref{pert} do not contain 
modes independent of $z$. This is related to the fact that 
Eqs.\eqref{ee1}--\eqref{ee6} no longer decouple into 
independent subsystems and all of the 6 radial field amplitudes 
$h^{(s)+}_{a}$, $\chi^{(s)}_{2}$, $\xi^{(s)}_{4}$  
are now non-vanishing.
Terms with  different $z$-dependence 
in \eqref{pert} now turn out to be coupled to each other, which is 
the consequence of the fact that the background fields
\eqref{ansatz1} depend explicitly on $z$.
 For example, it is no longer possible to have, as in Eq.\eqref{s1}, 
a solution with $h^{(1)+}_{2}\neq0$ but with $h^{(1)-}_{2}=0$, and so the analog of the 
solution \eqref{homogen} will read 
$$
\delta\Phi_2={A}\,\Psi_0(\rho)e^{Kt}+B({\cal I},t,\rho)e^{\pm 2i{K}z},
$$
where $B({\cal I},t,\rho)\neq 0$ if ${\cal I}\neq 0$. 
Since all the unstable modes depend on $z$,
the homogeneous negative mode is therefore {\it absent} 
for twisted strings  

Let us consider the effect of the perturbations on the gauge invariant 
vortex current.  With Eq.\eqref{cur} one obtains 
\be
\delta J_z=\Re(i\delta\Phi_2^\ast\, \partial_z\Phi_2+i\Phi_a^\ast\,\partial_z\delta\Phi_2+
2A_z\Phi_2^\ast\delta\Phi_2+f_2^2\delta A_z)
\ee
and using Eqs.\eqref{ansatz1},\eqref{pert} gives the growing part of this expression in the form 
\be                                    \label{Jz}
\delta J_z =f_2(\rho)\sum_{\kappa=-2K}^{2K}e^{\omega  t}
({\cal A}_\kappa(\rho)\cos(\kappa z)+{\cal B}_\kappa(\rho)\sin(\kappa z)),
\ee
where $\omega=\sqrt{|\omega_{(1)}^2|}$ for $\kappa<0$ and 
$\omega=\sqrt{|\omega_{(2)}^2|}$ for $\kappa>0$. 
A similar expression can be obtained for $\delta J_0$. 
Every unstable mode with $\kappa\in (-2K,2K)$
therefore produces ripples with the wavelength $\lambda=2\pi/|\kappa|$ 
on the homogeneous distribution
of the background current density. 
As the amplitude of these ripples grows in time, 
the current deforms more and more thus  tending to evolve  into a non-uniform `sausage like'
structure characterized by zones of charge accumulation.

\begin{figure}[ht]
\hbox to\linewidth{\hss%
	
	\resizebox{12cm}{4cm}{\includegraphics{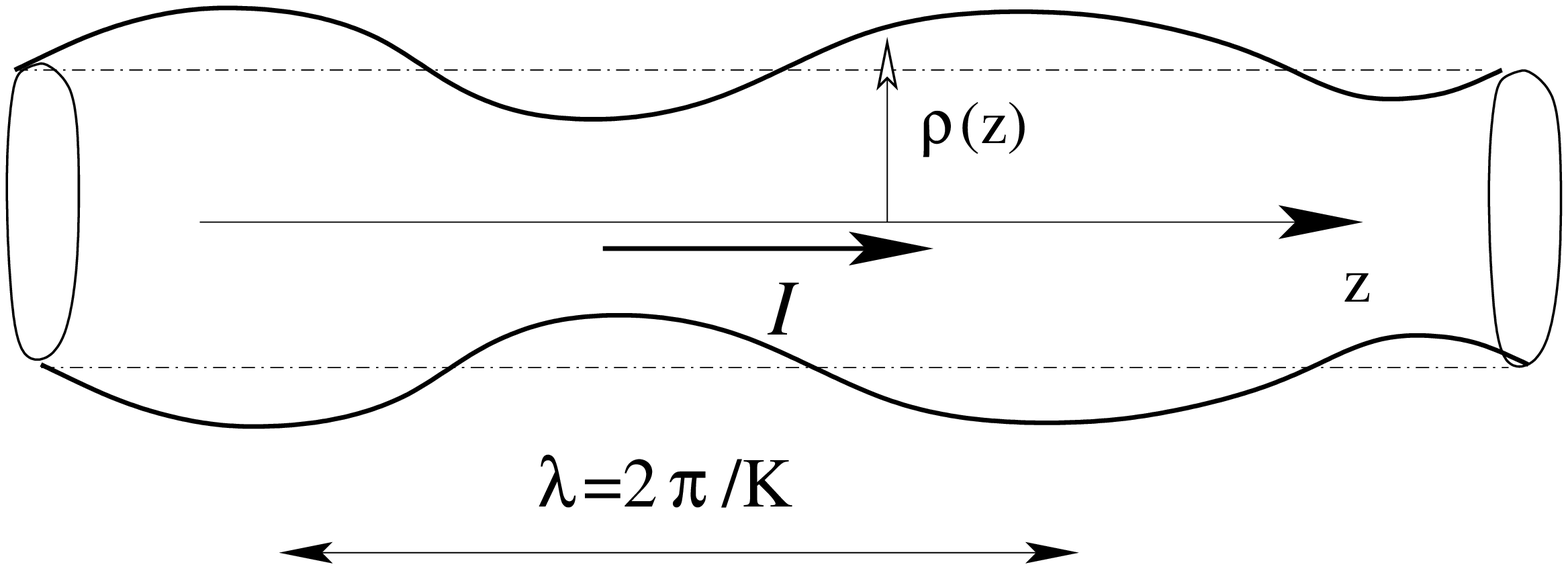}}
\hss}
\caption{Schematic view of the perturbed vortex current.
}
\label{Fig}
\end{figure}

Since the instability exists only for a finite range of $\kappa$,
there is a minimal instability wavelength corresponding to the maximal value of $\kappa$, 
\be
\lambda_{\rm min}=\frac{\pi}{K}\,.
\ee
As a result, finite vortex pieces obtained
by imposing  periodic boundary conditions  on the infinite vortex will 
have no room for inhomogeneous instabilities
if their length $L$ is less than $\lambda_{\rm max}$.
Next, as we already know, twisted vortices do not have homogeneous negative modes,
which means modes periodic with any period (for example, the $\kappa=0$ mode in 
\eqref{Jz} is homogeneous but  not negative).  
It follows then that short vortex pieces do not have negative modes at all.
They are therefore {\it stable}.

At first view the condition $L<\pi/K$ is impossible, since the background 
vortex configuration \eqref{ansatz1} 
contains $\exp(iKz)$ and so $L$ should be an integer multiple of $2\pi/K$. 
%This is true indeed in the context of the original semilocal model. The situation however changes 
However, 
if the semilocal model is viewed as a subsector of the local 
$\mathbf{SU}(2)\times \mathbf{U}(1)$ theory  
%theory 
in the limit where the $\mathbf{SU}(2)$ gauge field decouples, 
it is  possible to use the 
local 
%$\mathbf{SU}(2)\times \mathbf{U}(1)$
gauge transformations \eqref{loc} not merely as an intermediate 
technical tool as was done above, 
but in order
to pass to and stay in a physically admissible gauge where the 
vortex fields do not depend on $z$.
For example, applying to \eqref{ansatz1} the gauge 
transformation \eqref{loc} generated by  
\be
\mathbf{U}=\left(
\begin{array}{cc}
    1 & 0 \\
   0  & e^{-\imat K z}
\end{array}\right)       \nonumber 
\ee
gives (setting for simplicity $\Omega=M=0$) 
\begin{align}
A^{[2]}_\mu d x^\mu=K(a_2(\rho)-\frac12)\,dz+Na_1(\rho) d\varphi,~~~~~
W^{[2]}_\mu d x^\mu= \frac{\tau^3}{2}K dz,~~~~~
  \PHI^{[2]} =\left(\begin{array}{c}f_1(\rho) \EXP{\imat N\varphi} \\
    f_2(\rho)  \end{array}\right).        \nonumber 
\end{align}
This gauge is globally regular and contains no explicit $z$ dependence. One can therefore 
always work in this gauge and identify $z$ with any period.

To recapitulate, short pieces of twisted $N=1$ vortices are stable, but they 
become unstable with respect 
to fragmentation if their length exceeds a certain critical value. It 
is worth mentioning an interesting
analogy with the well known  Plateau-Rayleigh instability of a fluid 
cylinder in hydrodynamics
(see e.g. \cite{Plateau}). 
The cylinder is stable if only it is short enough, but when it gets 
longer the surface tension tends to 
break it up into disjoint pieces.  This can easily be seen when water 
comes out of a kitchen tap: 
while close to the tap the water jet is homogeneous, far enough from 
it the surface ripples appear.  
This hydrodynamical analogy seems to be not accidental, since the 
twisted superconducting vortex 
can effectively be described as a superposition of two charged 
fluids flowing in the opposite directions
\cite{SL}. Another interesting analogy that could be mentioned 
comes from a completely different domain: 
the Gregory-Laflamme instability of black strings in the 
gravity theory in higher dimensions \cite{GL}.
Black strings become unstable with respect to inhomogeneous 
perturbations if their length exceeds 
a certain critical value, which  can be explained by the 
tendency of the event horizon to minimize its area thus 
reducing the black string entropy.  

It is worth mentioning that 
in both of these two examples, and also in our case, the unstable modes exist
only in the axially symmetric $m=0$ sector.

\section{Concluding remarks}

Summarizing what has been said above, we considered generic field fluctuations around the 
twisted current carrying vortices in the 
$\mathbf{SU}(2)_{\rm global}\times \mathbf{U}(1)_{\rm local }$ semilocal field model,
or equally in the electroweak theory in the limit where the weak mixing angle is $\pi/2$. 
By studying the negative modes we have concluded that  
the twisted vortices exhibit essentially the same instabilities as the embedded ANO vortices, 
but with one important exception: they do not have the uniform spreading instability. 
All the negative modes of the fundamental $N=1$ twisted vortex are non-uniform, 
with the  wavelength bounded from below by 
$\pi/|K|$, where $K$ is the vortex twist parameter. 
Since the uniform instability is absent, it follows that  
short enough vortex pieces should be stable, since they have no room to 
accommodate the  non-uniform longwave instabilities. Longer vortex pieces 
will be unstable and the instability  will lead to their fragmentation 
into non-uniform objects characterized by 
zones of charge accumulation and by a non-uniform current density. 
However,  to fully describe this process 
requires going beyond the linearized approximation. 

It is interesting to note that increasing the vortex current ${\cal I}$ decreases the twist $K$
(see Fig.\ref{Fig1}) and so that the minimal instability wavelength $\pi/|K|$ increases,
which makes stable longer and longer vortex pieces. In addition, the characteristic 
time of the instability growth, $2\pi/|\omega|$, also increases, since $|\omega|<|K|$ 
(see Fig.\ref{Fig5}). As a result, increasing the current removes 
more and more of negative modes thus making the vortex `more and more stable'.

The non-uniform character of perturbations suggests that if there is a stable configuration 
to which the perturbed vortex could finally relax, 
then this configuration should be inhomogeneous. 
The current should be homogeneous in such a state, though. 
 Specifically,
integrating the current conservation condition $\partial_\mu J^\mu=0$ over a 3-volume sandwiched
between $z=z_1$ and $z=z_2$ planes  gives
$$
\frac{dQ}{dt}={\cal I}(z_1)- {\cal I}(z_2),
$$
where $Q$ is the total charge in the volume and ${\cal I}(z_1)$ 
is the total current through the 
$z=z_1$ plane, similarly for ${\cal I}(z_2)$. The condition ${\cal I}(z_1)\neq {\cal I}(z_2)$
therefore requires that $\frac{dQ}{dt}\neq 0$, and so 
inhomogeneities of the current can only appear during the dynamical phase 
when the time derivatives are non-zero.  
However, if the perturbed vortex finally ends up in an  
equilibrium state which is stationary, then one will have ${\cal I}(z_1)={\cal I}(z_2)$
and so the current will be homogeneous. At the same time, as we already know, 
the vortex itself cannot be homogeneous in a stable state (since the 
homogeneous vortex is unstable). 
We therefore conclude that if a stable stationary state of a  
current carrying vortex exists, then in this state the current 
will be constant along the vortex, 
but the vortex configuration itself will depend non-trivially on $z$ 
and perhaps also on $\varphi$. 
An interesting problem would be to look for such a stable state via
minimizing the energy by keeping the current and also total charge  fixed.   

Another possibility to have  stable objects would be to take short vortex pieces
and to close them to make small loops.  
If  exist as stationary field theory solutions, they might be stable. 
The issue of constructing such solutions is actually  quite important, 
since stationary vortex loops stabilized by centrifugal force
(they are called vortons \cite{vortons}) 
have been extensively discussed for about last 20 years 
(see \cite{vil} for a review). However, to the best of our knowledge, such objects have 
never been constructed by resolving the field equations of the underlying  field theory,
excepting  two examples obtained in a {\it rigid} and not local gauge field theory \cite{vort}.

It is finally worth noting that the above results are likely to be generalizable 
in the context of the full electroweak theory for arbitrary values of the weak mixing angle, 
since the analogs of the twisted vortices are known to exist in this case \cite{MV}. 
It seems plausible that short pieces of these superconducting electroweak vortices
could also be stable, and also perhaps small vortex loops.

\section*{Acknowledgements}
We thank Mark Hindmarsh for interesting discussions. 
This work was partly supported by the ANR grant NT05-$1_{-}$42856  `Knots and Vortons'.

%\newpage
%\appendix
%\section{\label{appendiceA}Equation of the fluctuations}

\section*{Appendix}

\renewcommand{\theequation}{A.\arabic{equation}}
\setcounter{equation}{0}

We derive here the full system of the perturbation equations. 
The starting point is the 
mode decomposition \eqref{fluct}, inserting which to the fluctuation equations \eqref{lin} 
and separating the $t,z,\varphi$ variables gives, for each set 
of values of $\omega$, $\kappa$, $m$,  
a system of 16 ODE's
for the  16 radial field amplitudes in \eqref{fluct}. 
These equations split into two independent subsystems of 8 equations each 
which have exactly the same structure, upon the identification \eqref{relation1}. 
Equations of the first group involve the amplitudes $\phi_a,\nu_a,\xi_1,\chi_2,\chi_3,\chi_4$:
\begin{align}
E_1:=&\left[-\D_{+}+\frac{N^2 (a_1-1)^2+m^2}{\rho^2}
+\beta (3 f_1^2+f_2^2-1)+(K^2-\Omega^2) a_2^2
+\kappa^2-\omega^2\right] \phi_1 \notag \\
+&2 \beta f_1 f_2 \,\phi_2
+2 \left((\Omega \omega-K \kappa) a_2+\frac{N m (1-a_1)}{\rho^2}\right) \nu_1
-2 \Omega a_2 f_1 \xi_1
+2 K a_2 f_1\, \xi_3 \notag \\
+&\frac{2 N f_1 (a_1-1)}{\rho^2 }\,\xi_4=0,                                 \label{E1}   
\end{align}
\begin{align}
E_2:=&\left[-\D_{+}+\frac{N^2 (1-a_1)^2+m^2}{\rho^2}
+\beta (f_1^2+f_2^2-1)+(K^2-\Omega^2) a_2^2
+\kappa^2-\omega^2\right] \nu_1 \notag \\
+&2 \left((\Omega \omega-K \kappa) a_2+\frac{N m (1-a_1)}{\rho^2}\right) \phi_1
+\omega f_1\, \xi_1
+f_1 \chi_2^\prime+\left(2 f_1^\prime+\frac{f_1}{\rho}\right) \chi_2 \notag \\
-&\kappa f_1\, \xi_3-\frac{m f_1}{\rho^2}\, \xi_4=0,                          \label{E2} 
\end{align}
\begin{align}
E_3:=&\left[-\D_{+}+\frac{(N a_1-M)^2+m^2}{\rho^2}
+\beta (f_1^2+3 f_2^2-1)+(K^2-\Omega^2) (a_2-1)^2
+\kappa^2-\omega^2\right] \phi_2 \notag \\
+&2 \beta f_1 f_2\, \phi_1
+2 \left((\Omega \omega-K \kappa) (a_2-1)+\frac{m (M-N a_1)}{\rho^2}\right) \nu_2
+2 \Omega f_2 (1-a_2)\, \xi_1  \notag \\
+&2 K f_2 (a_2-1)\, \xi_3
+\frac{2 f_2 (N a_1-M)}{\rho^2}\, \xi_4=0,                            \label{E3}    
\end{align}
\begin{align}
E_4:=&\left[-\D_{+}+\frac{(N a_1-M)^2+m^2}{\rho^2}+(K^2-\Omega^2) (a_2-1)^2
+\beta (f_1^2+f_2^2-1)
+\kappa^2-\omega^2\right] \nu_2 \notag \\
+&2 \left((\Omega \omega-K \kappa) (a_2-1)+\frac{m (M-N a_1)}{\rho^2}\right) \phi_2
+\omega f_2\, \xi_1
+f_2 \chi_2^\prime+\left(2 f_2^\prime+\frac{f_2}{\rho}\right) \chi_2 \notag \\
-&\kappa f_2\, \xi_3
-\frac{m f_2}{\rho^2}\, \xi_4=0,                                        \label{E4} 
\end{align}
\begin{align}
E_5:=&\left[-\D_{+}+\frac{m^2}{\rho^2}+2 (f_1^2+f_2^2)-\omega^2\right] \xi_3
+4 K f_1 a_2 \phi_1
+4 K f_2 (a_2-1) \phi_2 \notag \\
-&2 \kappa f_1\, \nu_1
-2 \kappa f_2\, \nu_2
+\omega \kappa\, \xi_1+
\kappa \left(\chi_2^\prime+\frac{\chi_2}{\rho}\right)
-\frac{\kappa m}{\rho^2}\, \xi_4=0,                                    \label{E5}         
\end{align}
\begin{align}
E_6:=&\left[-\D_{-}+2 (f_1^2+f_2^2)+\kappa^2-\omega^2\right] \xi_4+
4 f_1 N (a_1-1) \phi_1+
4 f_2 (N a_1-M) \phi_2 \notag \\
-&2 m f_1\, \nu_1
-2 m f_2\, \nu_2
+\omega m\, \xi_1
+m \left(\chi_2^\prime-\frac{\chi_2}{\rho}\right)
-\kappa m\, \xi_3 =0,                                                  \label{E6}   
\end{align}
\begin{align}
E_7:=&\left[-\D_{+}+\frac{m^2}{\rho^2}+2 (f_1^2+f_2^2)+\kappa^2\right] \xi_1
+4 \Omega f_1 a_2\, \phi_1
+4 \Omega f_2 (a_2-1) \phi_2  \notag \\
-&2 \omega f_1 \,\nu_1
-2 \omega f_2 \,\nu_2
+\omega \left(\chi_2^\prime+\frac{\chi_2}{\rho}\right)
-\omega \kappa\, \xi_3
-\frac{\omega m}{\rho^2}\, \xi_4=0,                                     \label{E7}    
\end{align}
\begin{align}
E_8:=&\left(\frac{m^2}{\rho^2}+2 (f_1^2+f_2^2)+\kappa^2-\omega^2\right) \chi_2 
+2 (f_1^\prime \nu_1-f_1 \nu_1^\prime)
+2 (f_2^\prime  \nu_2-f_2 \nu_2^\prime)
+\omega \xi_1^\prime
 \notag \\
-&\kappa \xi_3^\prime  
-\frac{m \xi_4^\prime}{\rho^2}=0.                                               \label{E8}
\end{align}
Here $\D_\pm:=\frac{d^2}{d\rho^2}\pm\frac{1}{\rho}\frac{d}{d\rho}$,
the functions $f_1,f_2,a_1,a_2$ and the constants $\Omega,K,N,M$ refer to the background 
twisted vortex solution.

Not all of these equations are independent.  Using the background field equations \eqref{4ode} 
one can check that the following identity holds,
\begin{equation}                              \label{id}
E_8^\prime+\frac{E_8}{\rho}-\frac{m}{\rho^2}\,E_6-\kappa E_5+\omega E_7-2f_1E_2-2f_2E_4=0,
\end{equation}
and so one of the equations is redundant.

A good consistency check for the equations is provided by the translational modes. 
If $A_\mu(x^\alpha),\PHI(x^\alpha)$ is the background  vortex solution written 
in the gauge \eqref{ansatz1}, then $A_\mu(x^\alpha+\X^\alpha),\PHI(x^\alpha+\X^\alpha)$,
where $\X^\alpha$ is a constant vector, will also be a solution. 
This implies that the Lie derivatives of $A_\mu(x^\alpha),\PHI(x^\alpha)$
 along $\X^\alpha$, 
\be                          \label{Lie}
\delta A_\mu=\X^\alpha\partial_\alpha A_\mu+A_\alpha\partial_\mu\X^\alpha,~~~
\delta \PHI=\X^\alpha\partial_\alpha \PHI,~~~
\ee
should fulfill the linearized field equations. If $\X=\frac{\partial}{\partial x}$, the vector 
generating displacements of the vortex in the $x$ direction, then calculating 
the Lie derivatives \eqref{Lie} and transforming them  according to 
\eqref{old} gives the perturbations in the form \eqref{fluct} with  $m=1$, $\omega=\kappa=0$ and 
\begin{align}
\phi_a&=f_a^\prime,~~~~\nu_1=-\frac{N}{\rho}\,f_1,~~~~\nu_2=-\frac{M}{\rho}\,f_2,~~\notag \\
\xi_1&=\Omega a_2^\prime,~~~\chi_2=\frac{N}{\rho^2}\,a_1,~~~\xi_3=Ka_2^\prime,~~~
\xi_4=N(a_1^\prime-\frac{1}{\rho}\,a_2). 
\end{align}
Using the background field equations \eqref{4ode} one can check that these fulfill 
Eqs.\eqref{E1}--\eqref{E8}. Choosing similarly $\X=\frac{\partial}{\partial z}$ gives 
$\omega=\kappa=m=0$ and 
\begin{align}
\phi_a&=0,~~~~\nu_1=0,~~~~\nu_2=f_2,~~~~\xi_1=\chi_2=\xi_3=\xi_4=0,
\end{align}
which also solves the equations.  

Using again Eqs.\eqref{4ode} one can check that 
Eqs.\eqref{E1}--\eqref{E8} 
are invariant under the gauge transformations \eqref{alpha},
\be                      \label{A10}
\phi_a\to\phi_a,~~\nu_a\to\nu_a+f_a\gamma,~~\xi_1\to\xi_1+\omega\gamma,~~
\chi_2\to\chi_2+\gamma^\prime,~~\xi_3\to\xi_3+\kappa\gamma,~~
\xi_4\to\xi_4+m\gamma\,,
\ee
where $\gamma=\gamma(\rho)$. 
In order to fix this gauge symmetry  we impose the temporal gauge condition $\delta A_0=0$
by setting 
$\xi_1=0$. 
We  also specialize to the case of purely magnetic backgrounds,
$
\Omega=0.
$
Eq.\eqref{E7} can then be resolved with respect to $\xi_3$ (see Eq.\eqref{xi4}). 
At the same time one can exclude from consideration Eq.\eqref{E5}
containing derivatives of $\xi_3$, since  we know that 
one of the equations in the system is redundant. 
Inserting then \eqref{xi4} into the remaining equations 
\eqref{E1}--\eqref{E4},\eqref{E6},\eqref{E8}
gives  the 6 independent second order equations 
for 6 field amplitudes $\phi_a$, $\nu_a$, $\chi_2$, $\xi_4$,
\begin{align}
&\left[-\D_{+}+\frac{N^2 (a_1-1)^2+m^2}{\rho^2}
+\beta (3 f_1^2+f_2^2-1)+K^2 a_2^2
+\kappa^2-\omega^2\right] \phi_1  
+2 \beta f_1 f_2 \phi_2                                 \notag \\
&+2 \left(\frac{N m (1-a_1)}{\rho^2}
-\kappa K a_2
-\frac{2 K a_2 f_1^2}{\kappa}\right) \nu_1
-\frac{4 K a_2 f_1 f_2 }{\kappa}\,\nu_2  
+\frac{2 K f_1 a_2}{\kappa} \left(
\chi_2^\prime+\frac{\chi_2}{\rho}\right)   \notag \\
&+\frac{2 f_1}{\rho^2} \left(N (a_1-1)
-\frac{K m a_2}{\kappa}\right) \xi_4 =0 \,,                \label{ee1} 
\end{align}
\begin{align}
&\left[-\D_{+}+\frac{N^2 (1-a_1)^2+m^2}{\rho^2}+\beta (f_1^2+f_2^2-1)
+K^2 a_2^2+2 f_1^2+\kappa^2-\omega^2\right] \nu_1 \notag \\
&+2 \left(\frac{N m (1-a_1)}{\rho^2}-K\kappa a_2 \right) \phi_1
+2 f_1 f_2\, \nu_2  +2 f_1^\prime\, \chi_2=0\,,             \label{ee2}
\end{align}
\begin{align}           
&\left[-\D_{+}+\frac{(N a_1-M)^2+m^2}{\rho^2}
+\beta (f_1^2+3 f_2^2-1)+K^2 (a_2-1)^2
+\kappa^2-\omega^2\right] \phi_2  
+2 \beta f_1f_2\, \phi_1                                   \notag \\
&+\frac{4 K (1-a_2)f_1f_2}{\kappa}\, \nu_1
+2 \left(\frac{m (M-N a_1)}{\rho^2}+K \kappa (1-a_2)
+\frac{2 K (1-a_2) f_2^2}{\kappa}\right) \nu_2 \notag \\
&+\frac{2 Kf_2 (a_2-1)}{\kappa} 
\left(\chi_2^\prime+\frac{\chi_2}{\rho}\right)
+\frac{2 f_2}{\rho^2} \left(N a_1-M
+\frac{K m(1-a_2)}{\kappa}\right) \xi_4=0\,,                       \label{ee3}
\end{align}
\begin{align}
&\left[-\D_{+}+(
\frac{(M-N a_1)^2+m^2}{\rho^2}+\beta (f_1^2+f_2^2-1)
+K^2 (1-a_2)^2+2 f_2^2+\kappa^2-\omega^2\right] \nu_2 \notag \\
&+2 \left(\frac{m (M-N a_1)}{\rho^2}
+K \kappa (1-a_2)\right) \phi_2
+2 f_1 f_2\, \nu_1
+2 f_2^\prime\, \chi_2=0\,,                                           \label{ee4}
\end{align}
\begin{align}
&\left[-\D_{-}+\frac{m^2}{\rho^2}+2 f_2^2+2 f_1^2+\kappa^2-\omega^2\right] \xi_4 % \notag \\
&+4 N f_1 (a_1-1) \phi_1
+4 f_2 (N a_1-M) \phi_2
-\frac{2 m}{\rho}\, \chi_2=0\,,                                   \label{ee5}
\end{align}
\begin{align}
&\left[-\D_{+}+\frac{m^2+1}{\rho^2}+2 f_1^2+2 f_2^2+\kappa^2-\omega^2\right] \chi_2
+4 f_1^\prime\, \nu_1+4f_2^\prime\, \nu_2
-\frac{2 m}{\rho^2}\, \xi_4=0.           \label{ee6}
\end{align}
As explained in the main text, there is no gauge freedom left in this system. 
These equations are invariant under 
\be                                      \label{minus}
m\to -m,~~~\kappa\to -\kappa,~~~\omega^2\to\omega^2,~~~\phi_a\to\phi_a,~~~
\nu_a\to -\nu_a,~~~\xi_4\to\xi_4,~~~\chi_2\to -\chi_2\,.
\ee
Taking suitable linear combinations of the 6 field amplitudes  
$\phi_a$, $\nu_a$, $\chi_2$, $\xi_4$ 
and their first derivatives one can build  a 6-component vector $\Psi$ using which 
the 6 equations \eqref{ee1}--\eqref{ee6} 
can be rewritten in the form of a 
Schrodinger-type eigenvalue problem \eqref{eig}.

In the ANO limit 
Eqs.\eqref{ee1}--\eqref{ee6} split into two 
independent subsystems. The first subsystem is obtained by setting in 
Eqs.\eqref{ee1},\eqref{ee2},\eqref{ee5},\eqref{ee6} $f_2=a_2=0$: 
\begin{align}
&\left[-\D_{+}+\frac{N^2 (a_1-1)^2+m^2}{\rho^2}
+\beta (3 f_1^2-1)
+\kappa^2-\omega^2\right] \phi_1 
+\frac{2N (1-a_1)}{\rho^2}\,(m\nu_1-f_1\xi_4)=0\,,              \notag  \\
&\left[-\D_{+}+\frac{N^2 (1-a_1)^2+m^2}{\rho^2}+\beta (f_1^2-1)
+2 f_1^2+\kappa^2-\omega^2\right] \nu_1 
+\frac{2N m (1-a_1)}{\rho^2}\, \phi_1
+2 f_1^\prime\, \chi_2=0\,,                                     \notag  \\  
&\left[-\D_{-}+\frac{m^2}{\rho^2}+2 f_1^2+\kappa^2-\omega^2\right] \xi_4  
+4 N f_1 (a_1-1) \phi_1
-\frac{2 m}{\rho}\, \chi_2=0\,,                                  \notag  \\
&\left[-\D_{+}+\frac{m^2+1}{\rho^2}+2 f_1^2+\kappa^2-\omega^2\right] \chi_2
+4 f_1^\prime\, \nu_1
-\frac{2 m}{\rho^2}\, \xi_4=0.                          \label{NO}
\end{align}
Since these equations 
do not contain perturbations of the second component of the Higgs field,
they actually describe the dynamics of the perturbed ANO vortex
within the original ANO model.

Eqs.\eqref{ee3},\eqref{ee4} in the ANO limit reduce to two decoupled
equations for the perturbations of the second component of the Higgs field
$h^{\pm}_2=\phi_2\pm \nu_2$
\begin{align}                                        \label{SS} 
&\left[-\D_{+}+\frac{(N a_1-M\mp m)^2}{\rho^2}
+\beta (f_1^2-1)+(K\pm\kappa)^2-\omega^2\right] h^\pm_2=0.  
\end{align}


\begin{thebibliography}{9}


\bibitem{ANO}  A.A.~Abrikosov, {\sl Sov.Phys. JETP} {\bf 5}  (1957) 1174;
 H.B.~Nielsen, P.~Olesen, {\sl Nucl.Phys.} {\bf B 61} (1973) 45.

\bibitem{vil}
A.~Vilenkin and E.P.S.~Shellard, {\it Cosmic Strings and Other Topological Defects.}
C.U.P. Cambridge (1994); M.~Hindmarsh and T.~Kibble, 
{\sl Rep.Prog.Phys.} {\bf 58}  (1995) 477.

\bibitem{sigrist}
M.~Sigrist and K.~Ueda, 
{\sl Rev.Mod.Phys.} {\bf 63}  (1991) 239.

\bibitem{vol}
M.M.~Salomaa and G.V.~Volovik, 
{\sl Rev.Mod.Phys.} {\bf 59}  (1987) 533.



\bibitem{achuc}
A.~Achucarro, T.~Vachaspati, 
{\sl Phys.Rep.} {\bf 327}  (2000) 427.

%\bibitem{baba}
%J.S.~Smorgrav, J.~Smiseth, E.~Babaev and A.~Sudbo,
%{\sl Phys.Rev.Lett.} {\bf D 94}  (2005) 096401. 

\bibitem{baba}
G.E.~Volvok, {\it The Universe in a helium droplet.}
Int.Ser.Monogr.Phys.  {\bf 117}  (2006);\\
S.~Smorgrav, J.~Smiseth, E.~Babaev and A.~Sudbo,
{\sl Phys.Rev.Lett.} {\bf D 94}  (2005) 096401. 



\bibitem{Shifman}
M.~Shifman and A.~Yung, 
{\sl Phys.Rev.} {\bf D 70}  (2004) 045004;\\
D.~Tong, {\tt hep-th/0509216}; 
A.~Hanany and Tong, {\sl JHEP} {\bf 0307}  (2003) 037. 

\bibitem{Witten}
E.~Witten, 
{\sl Nucl.Phys.} {\bf B 249}  (1985) 557.

\bibitem{vacha}
 T.~Vachaspati and  A.~Achucarro,
{\sl Phys.Rev.} {\bf D 44}  (1991) 3067. 

\bibitem{Hind}
M.~Hindmarsh, 
{\sl Phys.Rev.Lett.} {\bf 68}  (1992) 1263;
{\sl Nucl. Phys.} {\bf B 392} (1993) 461.



\bibitem{Gibbons}
G.W. Gibbons, M.E. Ortiz, F. Ruiz Ruiz and T.M. Samols, 
{\sl Nucl. Phys.} {\bf  B 358} (1992) 127;\\
E. Abraham. {\sl  Nucl. Phys.} {\bf B 399} (1993) 197.


\bibitem{SL} P.~Forgacs, S.~Reuillon, M.S.~Volkov, 
{\sl Phys.Rev.Lett. } {\bf 96} (2006) 041601; 
{\sl Nucl.Phys.} {\bf B751} (2006) 390.  




\bibitem{jack}
I.M.~Gel'fand and S.V.~Fomin, {\it Calculus of Variations},
Englewood Cliffs N.J., Prentice-Hall (1963). 


\bibitem{BV}
E.B.~Bogomolny and A.I.~Vainshtein,
{\sl Sov.J.Nucl.Phys.} {\bf 23}  (1976) 588.

\bibitem{Good}
M.~Goodband and M.~Hindmarsh, 
{\sl Phys.Rev.} {\bf D 52}  (1995) 4621.

\bibitem{MV95} 
M.S.~Volkov, O.~Brodbeck, G.V.~Lavrelashvili and N.~Straumann, 
{\sl Phys.Lett.} {\bf B 349}  (1995) 438.




\bibitem{baacke} J.~Baacke, {\sl Z.Phys.} {\bf C 53}  (1992) 399;
H.~Hollemann, {\sl Phys.Lett.} {\bf B 338} (1994) 181. 

\bibitem{Amm} H.~Amann, P.~Quittner, {\sl J.Math.Phys.} {\bf 36}  (1995) 4553.



\bibitem{Plateau}
J.~Eggers,
{\sl Rev.Mod.Phys.} {\bf 69}  (1997) 865.


\bibitem{GL}
R.~Gregory and R.~Laflamme,
{\sl Phys.Rev.Lett.} {\bf 70}  (1993) 2837;\\
V.~Cardoso and O.J.~Dias, 
{\sl Phys.Rev.Lett.} {\bf 96}  (2006) 101601. 

\bibitem{vortons} R.L.~Davis and E.P.S.~Shellard,
{\sl Nucl.Phys. } {\bf B 323} (1989) 209. 

\bibitem{vort} R.A.~Battye, N.R.~Cooper and P.M.~Sutcliffe,
{\sl Phys.Rev.Lett. } {\bf 88} (2002) 080401;\\
Y.~Lemperiere and E.P.S.~Shellard,
{\it Phys.Rev.Lett. } {\bf 91} (2003) 141601.

\bibitem{MV} M.S.~Volkov, 
{\sl Phys.Lett. } {\bf B 644} (2007) 203. 




\end{thebibliography}
\end{document}